\documentclass[twocolumn]{aastex62}
\usepackage{color,comment}
\usepackage[normalem]{ulem}
\usepackage[caption=false]{subfig}
\usepackage{graphicx}
\usepackage{verbatim}
\usepackage{multirow}
\usepackage{float}
\usepackage{wrapfig}
\received{}
\revised{}
\accepted{}
\shorttitle{SFHs \& Stellar Masses from SED Modeling}
\shortauthors{S. Lower et al.}
\begin{document}

\title{How Well Can We Measure the Stellar Mass of a Galaxy: \\The Impact of the Assumed Star Formation History Model in SED Fitting
}

\author[0000-0003-4422-8595]{Sidney Lower}
\affil{Department of Astronomy, University of Florida, 211 Bryant Space Science Center, Gainesville, FL, 32611, USA}
\author[0000-0002-7064-4309]{Desika Narayanan}
\affil{Department of Astronomy, University of Florida, 211 Bryant Space Science Center, Gainesville, FL, 32611, USA}
\affil{University of Florida Informatics Institute, 432 Newell Drive, CISE Bldg E251 Gainesville, FL, 32611, US}
\affil{Cosmic Dawn Centre at the Niels Bohr Institue, University of Copenhagen and DTU-Space, Technical University of Denmark}
\author[0000-0001-6755-1315]{Joel Leja}
\affil{Center for Astrophysics $|$ Harvard \& Smithsonian, 60 Garden St. Cambridge, MA 02138, USA}
\affil{NSF Astronomy and Astrophysics Postdoctoral Fellow}
\author[0000-0002-9280-7594]{Benjamin D. Johnson}
\affil{Center for Astrophysics $|$ Harvard \& Smithsonian, 60 Garden St. Cambridge, MA 02138, USA}
\author[0000-0002-1590-8551]{Charlie Conroy}
\affil{Center for Astrophysics $|$ Harvard \& Smithsonian, 60 Garden St. Cambridge, MA 02138, USA}
\author[0000-0003-2842-9434]{Romeel Dav{\'{e}}}
\affil{Institute for Astronomy, Royal Observatory, University of Edinburgh, Edinburgh, EH9 3HJ, UK}
\affil{University of the Western Cape, Bellville, Cape Town 7535, South Africa}
\affil{South African Astronomical Observatory, Observatory, Cape Town 7925, South Africa}

\begin{abstract}
The primary method for inferring the stellar mass ($M_*$) of a galaxy is through spectral energy distribution (SED) modeling. However, the technique rests on assumptions such as the galaxy star formation history and dust attenuation law that can severely impact the accuracy of derived physical properties from SED modeling. Here, we examine the effect that the assumed star formation history (SFH) has on the stellar properties inferred from SED fitting by ground truthing them against mock observations of high-resolution cosmological hydrodynamic galaxy formation simulations. Classically, star formation histories are modeled with simplified parameterized functional forms, but these forms are unlikely to capture the true diversity of galaxy SFHs and may impose systematic biases with under-reported uncertainties on results. We demonstrate that flexible nonparametric star formation histories outperform traditional parametric forms in capturing variations in galaxy star formation histories, and as a result, lead to significantly improved stellar masses in SED fitting. We find a decrease in the average bias of $0.4$ dex with a delayed-$\tau$ model to a bias under $0.1$ dex for the nonparametric model, though this is heavily dependent on the choice of prior for the nonparametric model. Similarly, using nonparametric star formation histories in SED fitting result in increased accuracy in recovered galaxy star formation rates (SFRs) and stellar ages.

\end{abstract}

\section{Introduction} \label{sec:intro}

The ability to accurately infer the physical properties of galaxies is critical for our understanding of galaxy formation and evolution. Modeling the ultraviolet (UV) to infrared (IR) spectral energy distributions (SEDs) of galaxies is one of the main methodologies used to derive the physical properties of galaxies such as the stellar mass (M$_*$), star formation rate (SFR), and stellar age. These techniques, pioneered by \cite{1968ApJ...151..547T}, \cite{1971ApJS...22..445S}, and \cite{1972A&A....20..361F}, have seen an explosion of interest and activity as space-based missions such as Galaxy Evolution Explorer (GALEX) and the Hubble Space Telescope have opened up ultraviolet and optical wavelengths for galaxies near and far respectively. Similarly, advances in infrared and submillimeter detector technology have opened up infrared windows that provide  constraints for SED models that consider energy balance between UV/optical photons and thermal infrared emission from dust. 

The abundance of panchromatic data available has spurred the development of many SED modeling codes (e.g. {\sc cigale}; \citealt{2019A&A...622A.103B}, {\sc fast}; \citealt{2009ApJ...700..221K}; \citealt{2018ascl.soft03008K}, and {\sc magphys}; \citealt{da_cunha_simple_2008}) that were developed to estimate physical properties from observed broadband data. These codes rely on models describing the stellar populations, nebular emission, and dust content in the galaxy, along with an optimization method to fit the SED and return the resulting physical parameters

The basic components in an SED model include information about stellar populations -- the stellar initial mass function (IMF), stellar isochrones and spectral templates, and star formation history (SFH) -- along with emission from nebular regions and active galactic nuclei (AGN), dust emission, and attenuation from dust. The robustness of an SED model and our ability to accurately recover physical properties of a galaxy depend on our confidence in each model component to accurately capture the complexity of the many physical processes that occur in a galaxy \citep[see e.g.][for an in depth review]{conroy_modeling_2013}.

Despite the widespread use of SED modeling by the observational galaxy community, it remains difficult to establish the efficacy of the technique due to the many weakly-constrained components and relative lack of ground-truth (see \citealt{2015ApJ...808..101M} and reviews by \citealt{conroy_modeling_2013}; \citealt{walcher_fitting_2011}). Indeed, some efforts have emerged in recent years to provide such a ground-truth in the context of extensive comparisons between SED modeling codes \citep[e.g.][]{hunt_comprehensive_2019} or comparisons between input mock SEDs drawn from known physical proprieties and the output properties from SED modeling. Examples of the latter context range from testing on a library of SEDs from an empirical mock catalog \citep{2015ApJ...808..101M, leja_how_2019} to SED modeling of idealized galaxies \citep{hayward_should_2015} and galaxies from a cosmological simulation \citep{iyer_reconstruction_2017, katsianis_high_2020}. 

One of the more influential yet poorly constrained components of SED fitting is the assumed form of the star formation history (SFH) \citep[e.g.][]{ocvirk_2006, iyer_reconstruction_2017, carnall_how_2019}. The most common models for SFHs are parameterized by a simple functional form, and the parameters varied in the SED fit describe that functional form. Hereafter, we refer to these as "parametric" star formation histories. Examples include the $\tau$ and delayed-$\tau$ models, which model the SFH as exponentially declining with some characteristic $e$-folding time. Although these models describe the SFHs of galaxies in a closed box (i.e. isolated with no inflow of pristine gas) where gas forms stars with constant star formation efficiency, the restricted nature of the functional forms does not match the diversity of true galaxy SFHs (\citealt{1984ApJ...284..544G}; \citealt{1986A&A...161...89S};  \citealt{2009ApJS..184..100L}; \citealt{2013ApJ...770...63O}; \citealt{simha_parametrising_2014}; \citealt{diemer_log-normal_2017}). \cite{carnall_how_2019} has shown the cosmic SFR density (CSFRD) inferred from delayed-$\tau$ SFH fits to galaxies from the GAMA survey \citep{gama} are incompatible with the CSFRD predicted by the Universe Machine \citep{behroozi_2019_univmach}. \cite{leja_older_2019} found that the backwards-evolved stellar mass functions (SMFs) inferred from SED fits to galaxies from the 3D-HST survey using a delayed-$\tau$ SFH model are in tension with observed SMFs at z$=3$. The assumed SFH and associated priors can also strongly bias the inferred physical properties of galaxies \citep{simha_parametrising_2014, Acquaviva_2015, salmon_relation_2015, Ciesla_2017, iyer_reconstruction_2017, carnall_how_2019, curtis-lake_modelling_2020}. For instance, \cite{michalowski_stellar_2012} found that the assumed star formation history model had the largest impact out of all other SED model components on the stellar masses inferred for observations of sub-millimeter galaxies (SMGs). Similarly \cite{2020MNRAS.494.3828D} found that the average difference between the stellar masses predicted by the {\sc magphys} SED model and the true stellar masses for galaxies from the {\sc eagle} cosmological simulation was close to $0.5$ dex. This bias, attributed to the assumed SFH module, is consistent with the earlier results of \citet{michalowski_stellar_2012}.

Parametric SFHs with more flexibility have also been explored in the literature \citep[e.g.][]{Papovich_2011, simha_parametrising_2014, ciesla_sfr-_2017, carnall_2018}. One example is the log-normal parametrization, which has been shown to reproduce the evolution of the cosmic star formation rate history \citep{gladders_imacs_2013} and provide a reasonable match to the Illustris galaxy SFHs \citep{diemer_log-normal_2017}. However, the log-normal parameterization still suffers from similar stellar age biases as the simpler parametric forms \citep{Ciesla_2017, iyer_reconstruction_2017, carnall_how_2019, leja_how_2019}. This happens because the SFH is constrained to \textit{either} have recent star formation \textit{or} have a population of old stars \citep{leja_how_2019} -- a consequence of the inflexible mathematical form. Moreover, \cite{diemer_log-normal_2017} fit a log-normal SFH directly to the SFHs of the Illustris galaxies instead of via SED modeling where all galaxy properties must be inferred simulatenously, thus biases such as outshining \citep{papovich_stellar_2001} did not affect the fit.

An alternative to the parametric models described above are nonparametric forms. Nonparametric SFH models, defined as models that do not explicitly assume a functional form, have been shown to have the flexibility to reproduce the variation in SFHs seen in observations and galaxy formation simulations. Examples include models sampled from a diverse basis of analytical SFH models \citep{iyer_reconstruction_2017}, flexible piece-wise linear functions in time \citep{reichardt_recovering_2001, heavens_2004, tojeiro_recovering_2007, kelson_decoding_2014, leja_deriving_2017, morishita_massive_2019, leja_how_2019}, models drawn from simulations or semi-analytical models (SAMs) \citep{2008ApJ...686.1503B, pacifici_relative_2012, Pacifici_2016, 2017ApJS..233...13Z}, and models utilizing machine learning methods \citep{lovell_2019, iyer_nonparametric_2019}. Most early implementations of nonparametric SED fitting methods relied on spectroscopic data but models like those presented in \citep[e.g.][]{iyer_reconstruction_2017, leja_deriving_2017} have been shown to produce reasonable results with broadband data only.

While ground-truth tests of parametric SFHs have generally found poor agreement with both real and simulated SFHs, the opposite is typically true for nonparametric models. Focusing on results from broadband SEDs, \cite{iyer_nonparametric_2019} validated the reconstruction of SFHs using Gaussian processes with a sample of galaxies from the Santa Cruz SAMs and the {\sc mufasa} hydrodynamical cosmological simulation and found good agreement between the mock galaxy SFHs and the SFHs reconstructed from Gaussian processes. \cite{leja_older_2019} demonstrated that galaxies selected from the 3D-HST photometry catalog (\citealt{2014ApJS..214...24S}) were inferred systematically more massive and older when modeled with the nonparametric SFHs in {\sc prospector} compared to previously published results using parametric SFHs. However, extensive tests and ground-truthing the results from SED fitting have lacked either in ground-truth sample technique (as in the case of e.g. \cite{leja_how_2019} where the mock galaxy SFHs used to generate synthetic SEDs were relatively simple), sample size, or a combination of both (as in the case of e.g. \cite{hayward_should_2015} where just two scenarios of simplified idealized simulated galaxies were used).

Thus, in this paper, we advance our understanding of the efficacy of SED modeling by fitting nonparametric SFH models to mock SEDs generated from 3D radiative transfer on a large galaxy sample from a hydrodynamical cosmological simulation. First, we focus on the impact of the assumed SFH on the stellar masses and other galaxy properties inferred from SED fitting by isolating the SED fits from dust. Specifically, we fix the dust-to-stellar geometries in our radiative transfer calculations and match that geometry in our SED fits to effectively obtain the baseline uncertainties on the inferred galaxy properties that can be attributed to just the assumed form of the SFH model. This is distinguished from previous observationally based studies, where all components of the SED model are necessarily tested at once. In other words, we ignore the uncertainty from other model components to first understand the biases and uncertainties imposed by just the SFH model. While the main goal of this paper is to isolate the impact of galaxy SFHs on SED fitting techniques, specifically to ground-truth the efficacy of various SFH models, we then briefly generalize these results by re-running our dust radiative transfer using the intrinsic dust-stellar geometries and dust properties from the simulations.

The paper is organized as follows: in \S~\ref{sec:num_methods}, we describe our numerical methods (including cosmological simulations, radiative transfer, and SED fitting).  In \S \ref{sec:results} we describe the results of the SED fitting through comparisons to the true values from the simulations. In \S \ref{sec:disc} we discuss the results, the possible origins of fitting failures, and the inclusion of realistic dust in the mock SEDs. In \S \ref{sec:conclusion} we conclude and propose a pathway towards improving our dust models to better accommodate realistic galaxies in SED fitting.

\section{Numerical Methods}\label{sec:num_methods}
\subsection{Overview}

For this analysis, we fit the SEDs of galaxies from a cosmological hydrodynamical simulation to determine the robustness of stellar masses estimated from SED modeling. The simulated galaxy SEDs are generated with post-processing radiative transfer that propagates the intrinsic stellar SEDs through dust in the interstellar medium (ISM). 

We fit these mock SEDs using the Bayesian inference code {\sc prospector}. We "observe" the mock SEDs in the same broad-band filters for all galaxies and use the same models for stellar metallicity and dust attenuation when fitting, so that the only difference between the results shown below originate from differences in the assumed star formation history model in the SED fit. In the remainder of this section, we describe these methods in more detail.

\subsection{The {\sc simba} Galaxy Formation Model} 
\label{section:simba}

\begin{figure*}
    \centering
    \includegraphics[width = 0.9\textwidth]{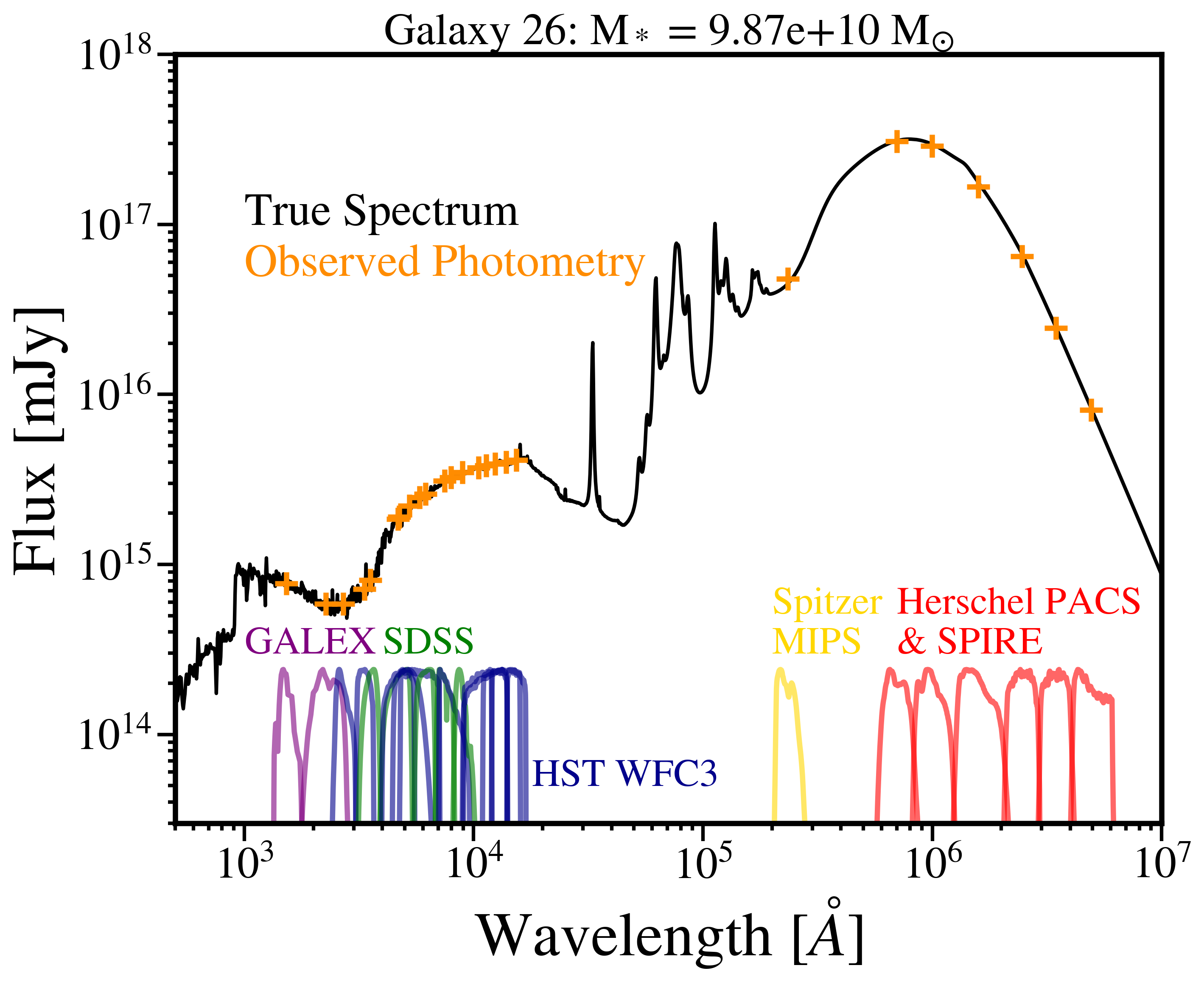}
    \caption{Example {\sc powderday} mock SED with 'observed' photometric bands highlighted, spanning from GALEX \textit{FUV} to \textit{Herschel} SPIRE, totaling $25$ bands resulting in almost complete coverage across all wavelength regimes. All galaxies are observed with the same filters, and photometric errors are fixed at $3\%$. Photometry is 'observed' for each filter and convolved over the filter bandwidth. We ignore NIR photometry, spanning from $\sim$ $2\mu$m to $20\mu$m, to avoid dependence on the ultra small grain size fractions chosen for the {\sc powderday} calculations.}
    \label{fig:SED}
\end{figure*}

We first need a population of model galaxies to generate the SEDs from.  For this, we use the {\sc simba} cosmological simulation, described in full in \cite{dave_simba:_2019}. Briefly, {\sc simba} is the descendant of the simulation suite {\sc mufasa} and relies on {\sc gizmo}'s meshless finite mass hydrodynamics. New sub-resolution prescriptions for stellar and AGN feedback as well as black hole growth have enabled {\sc simba} to accurately reproduce observables like the galaxy stellar mass function and the star forming main sequence. {\sc  simba} additionally includes an on-the-fly self-consistent model for the formation, growth, and destruction of dust that reproduces both the $z=0$ dust mass function, as well as the scaling between the dust to gas ratio and metallicity \citep{li_dust--gas_2019}. 

We employ a box with $25/h$ Mpc side length with $512^3$ particles, resulting in a baryon mass resolution of $1.4\times$ $10^6 \: \mathrm{M}_{\odot}$. To identify galaxies, we have employed a modified version of {\sc caesar}\footnote{https://github.com/dnarayanan/caesar} \citep{2014ascl.soft11001T}. We focus on the $z=0$ snapshot, in which there are $\sim 1600$ galaxies identified with a minimum of $32$ bound star particles with a 6D friends-of-friends galaxy finder. The choice of $32$ star particles is motivated by the mass resolution of this simulation. These galaxies lie within a mass range of $4.4 \times 10^7-1.4 \times 10^{12} M_\odot$.

\subsection{Dust Radiative Transfer}\label{sec:pd}

We use the 3D radiative transfer code {\sc powderday}\footnote{https://github.com/dnarayanan/powderday} \citep{pd_2020} to construct the synthetic SEDs by first generating with {\sc fsps} (\citealt{conroy_propagation_2009}; \citealt{conroy_propagation_2010}) the dust free SEDs for the star particles within each cell using the stellar ages and metallicities as returned from the cosmological simulations. For these, we assume a \cite{kroupa_initial_2002} stellar IMF and the {\sc mist} stellar isochrones \citep{2011ApJS..192....3P, 2016ApJS..222....8D, 2016ApJ...823..102C}. 

Traditionally, {\sc powderday} then propagates the emission from these stars through the diffuse dusty interstellar medium using {\sc hyperion} as the dust radiative transfer solver \citep{robitaille_hyperion:_2011,robitaille12a}. However, this then imposes the uncertainty of the diverse attenuation laws that vary from galaxy to galaxy on our SED fits \citep{narayanan18b,narayanan18a,2020arXiv200103181S}. We therefore abandon the diffuse dust in our {\sc powderday} radiative transfer simulations, and instead employ a dust screen surrounding all stars. This is akin to how {\sc prospector} treats dust obscuration, and therefore allows us to isolate the impact of the galaxy star formation history on our SED fits. In the dust screen setup for {\sc powderday}, we assume a uniform dust screen around all stars with an optical depth of $\tau_{\mathrm{uniform}}$ = $0.7$. Younger stars ($<$ $10$ Myr old) have an additional assumed source of extinction from their birth clouds that have an optical depth of $\tau_{\mathrm{BC}}$ = $0.7$. This fiducial dust screen model ensures an apples-to-apples comparison between the creation of the SEDs and the technique used to fit them. We then generalize this comparison in \S \ref{sec:diffuse_dust} where we re-run our dust radiative transfer with {\sc powderday} using the intrinsic dust-stellar geometries and dust properties from the simulations.

The result of the {\sc powderday} radiative transfer is the UV - FIR spectrum for each galaxy. We extract model photometry from these dust spectra, selecting 25 bands from the GALEX \textit{FUV} filter at $1542$ {\AA} through the \textit{Herschel} SPIRE band at $500$ $\mu$m as shown in Figure~\ref{fig:SED} and Table~\ref{table:filters}. The SED coverage is comparable to galaxies in the e.g. CANDELS GOOD-S field \citep{2013ApJS..207...24G}. And while SED coverage necessarily impacts the accuracy of galaxy properties inferred from SED fitting, we leave an in-depth investigation into these factors to a future work. Further, we fixed uncertainties to $3\%$ of the flux value, since the aim of this study is not to analyze the effect of photometric uncertainties but rather the systematics that arise from the use of various SFH models. 

\begin{table}[]
\begin{tabular}{llc}
\hline
Instrument     & Filter & Effective Wavelength (\AA) \\ \hline
GALEX          & \textit{FUV}    & 1549                      \\
               & \textit{NUV}    & 2304                      \\
\textit{HST}/WFC3       & F275W  & 2720                      \\
               & F336W  & 3359                      \\
               & F475W  & 4732                      \\
               & F555W  & 5234                      \\
               & F606W  & 5780                      \\
               & F814W  & 7977                      \\
               & F105W  & 10431                     \\
               & F110W  & 11203                     \\
               & F125W  & 12364                     \\
               & F140W  & 13735                     \\
               & F160W  & 15279                     \\
SDSS           & \textit{u}      & 3594                      \\
               & \textit{g}      & 4640                      \\
               & \textit{r}      & 6122                      \\
               & \textit{i}      & 7439                      \\
               & \textit{z}      & 8897                      \\
\textit{Spitzer}/MIPS   & 24 $\mu$m  & 232096                    \\
\textit{Herschel}/PACS  & Blue   & 689247                    \\
               & Green  & 979036                    \\
               & Red    & 1539451                   \\
\textit{Herschel}/SPIRE & PSW    & 2428393                   \\
               & PMW    & 3408992                   \\
               & PLW    & 4822635                   \\ \hline
\end{tabular}

\caption{Table of the 25 filters used to extract photometry from the synthetic {\sc powderday} spectra. Broadband fluxes are assigned a 3\% fractional uncertainty.}
\end{table}\label{table:filters}

\subsection{SED Fitting}\label{sec:sed_fitting}
%table for SED parameters
%recommend not expanding because it's pretty messy
\begin{table*}[]
\centering
\begin{tabular}{l|l|l|l}
                & Model                                 & Parameters                         & Prior Distribution                                                                                                                                                                                 \\ 
\hline\hline
                & Delayed $\tau$       & Age  & Uniform(0.01, 13.8) Gyr                                                                                                                                              \\
                &                                       & $\tau$            & LogUniform(0.001, 10) Gyr$^{-1}$                                                                                                                                \\
SFH             & Delayed $\tau$+burst & Age, $\tau$         & As above                                                                                                                                                                                           \\
                &                                       & Burst time as a fraction of age     & Uniform(0.5, 1.0) * Age                                                                                                                                              \\
                &                                       & Burst M$_*$ fraction            & Uniform(0.0, 5.0) * M$_*$                                                                                                                                       \\
                & Constant                              & Age                                & As above                                                                                                                                                                                           \\
                & Nonparametric                         & M$_*$ fraction per time bin (N = 10) & Dirichlet                           \\
                &                          & Concentration parameter $\alpha$ & Fixed at 0.7                                                                                                                                                                \\ 
\hline
M$_*$ - Z         & \citet{gallazzi_ages_2005}                   & Log(M$_*$ formed)               & Uniform(7, 13) M$_{\odot}$                                                                                                                       \\
                &                                       & Stellar Metallicity Z$_*$              & ClippedNormal(-1.9, 0.19)  \\ 
\hline
Dust Emission   & \citet{draine_infrared_2007}       & Minimum radiation field  U$_\mathrm{min}$          & Uniform(0.1, 30)                                                                                                                                                                                   \\
                &                                       & Warm dust fraction $\gamma$              & Uniform(0.0, 1.0)                                                                                                                                                                                  \\
                &                                       & PAH mass fraction  q$_\mathrm{PAH}$                & Fixed at 5.86\%                                                                                                                                                                                  \\ 
\hline
Dust Absorption & Fixed Dust Screen & Uniform dust screen opacity $\tau_\mathrm{uniform}$  & Fixed at 0.7 \\
        &        & Birth cloud dust opacity $\tau_\mathrm{BC}$ & Fixed at 0.7 \\
        &        & Powerlaw index & Fixed at -0.7 (uniform) \& -1.0 (birth cloud)
\end{tabular}
\caption{Table of SED models and associated parameters and prior distributions for {\sc prospector} SED fitting. All {\sc powderday} SEDs are fit four separate times with a different SFH model. Three commonly used parametric SFH models are considered (delayed-$\tau$, delayed-$\tau$ with a burst component, and a constant SFR across time) in addition to a nonparametric SFH model.  The dust and stellar mass$-$stellar metallicity models are kept the same between runs though the parameters are allowed to vary.}
\end{table*}\label{table:params}

In order to model the mock SEDs generated by {\sc powderday}, we use the Bayesian inference code {\sc prospector}\footnote{https://github.com/bd-j/prospector} \citep{leja_deriving_2017, leja_how_2019}. {\sc Prospector} derives galaxy physical properties using stellar population synthesis evolved with dust within the framework of {\sc fsps}. {\sc powderday} and {\sc prospector} rely on the same spectral libraries, IMF, and stellar isochrones within {\sc fsps}. And though the assumptions made when modeling stellar spectra, especially concerning the impact of late stellar evolutionary stages on a galaxy's SED, are important and the physical parameters derived from SED can be greatly influenced by these assumptions \citep[e.g.][]{2015ApJ...801...97S, 2015ApJ...808..101M}, this exploration is ultimately outside the scope of this paper as we are focusing on the targeted question of understanding the impact of the assumed SFH on derived stellar masses. Additionally, the use of the fiducial dust screen model enables the subsequent SED fitting procedures and results to be isolated from assumptions about the dust attenuation model. In other words, the models described below will vary due to assumptions about the SFH only, effectively isolating our results to the differences between these models and providing a baseline uncertainty estimate arising from just the SFH model. 

{\sc prospector} uses a Bayesian inference framework via {\sc dynesty} nested sampling \citep{2020MNRAS.tmp..280S} to fit the observed SEDs and provide posterior distribution functions (PDFs) for physical parameters such as stellar mass, stellar metallicity, and star formation rate. The power of {\sc dynesty} lies in its ability to efficiently sample multi-model distributions and have well-defined stopping criteria based on evaluations of Bayesian evidence ensuring model convergence. Below we describe the SED model components and their prior distributions, which are summarized in Table \ref{table:params}. Because {\sc dynesty} is based on Bayesian inference, all results in our analysis are sampled from the resulting PDF from the nested sampling iterations (as opposed to so-called 'best-fit' parameters quoted by a $\chi^2$ minimization algorithm), with uncertainties quoted as the $16^{th}$ through $84^{th}$ percentiles of the posterior distributions.

\subsubsection{Star Formation History}\label{sec:sfh}

\begin{figure*}
\centering
\includegraphics[width=0.48\textwidth]{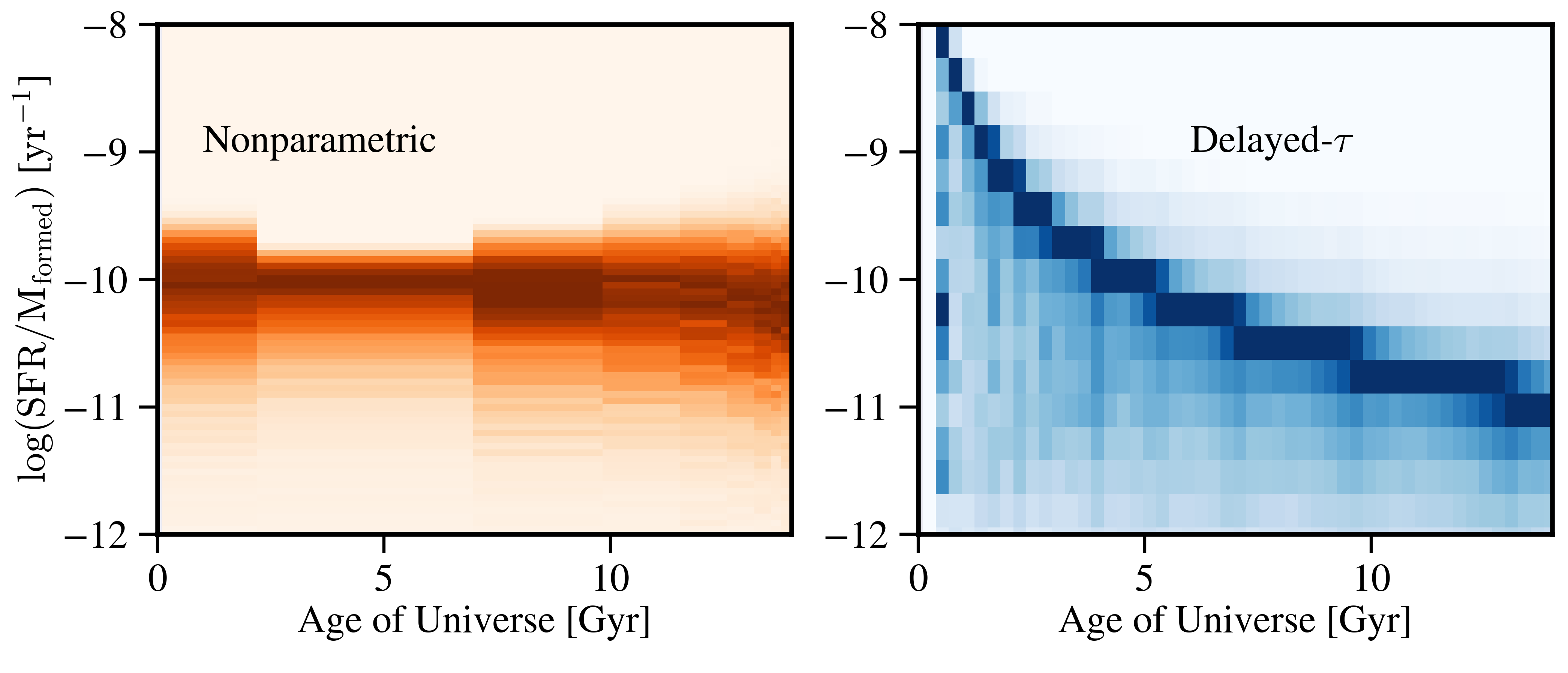}
\includegraphics[width=0.48\textwidth]{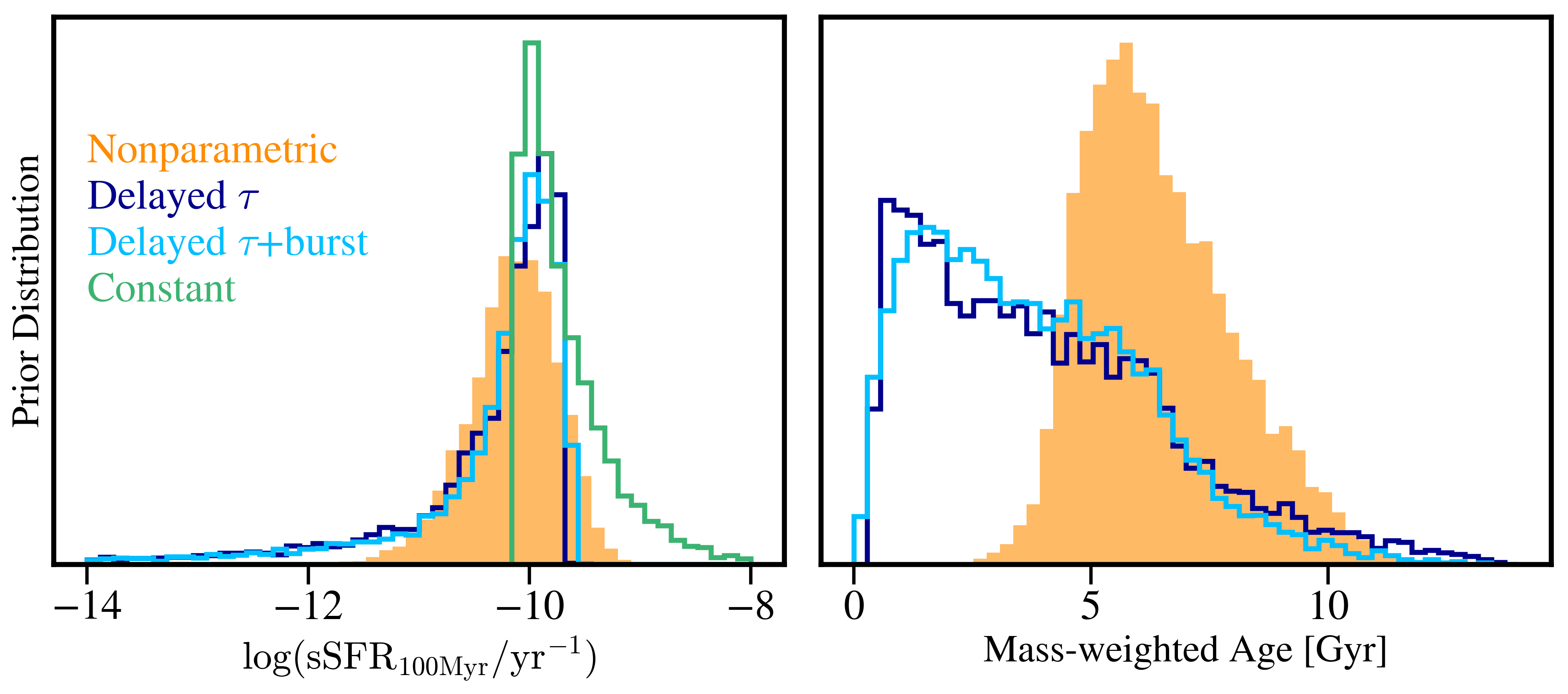}
\caption{\textbf{Left}: Prior probability distributions for the nonparametric SFH model and the Delayed-$\\tau$ model.  \textbf{Middle}: Effective prior distribution on the specific star formation rate averaged over the last 100 Myr. \textbf{Right}: Effective prior distribution on the mass-weighted stellar age. The constant SFH mass-weighted age prior is not computed as this is simply 0.5$\times t_H$.}
\label{fig:sfh_priors}
\end{figure*}

{\sc prospector} includes several models for a galaxy's SFH, including the commonly used delayed-$\tau$ model, along with flexible nonparametric models. The nonparametric models are constrained by priors that can either enforce continuity (i.e. the SFH is smooth rather than bursty) or that allow for episodes of SF bursts. We show the prior probability distribution for one such model in Figure~\ref{fig:sfh_priors}. As a means for comparison, we considered three simple parametric star formation histories, an example of which is also shown as a prior probability distribution in the left panel of Figure~\ref{fig:sfh_priors}. All star formation history models used in this analysis are described in depth in \citet{carnall_how_2019} and \citet{leja_how_2019}, while we provide a more top level view here. Three commonly used parametric models are considered (delayed-$\tau$, delayed-$\tau$ with a burst component, and a constant SFR across time) as well as a nonparametric SFH model.

\textbf{Delayed$-\tau$}: The delayed-$\tau$ model is an exponentially declining SFH, parameterized by the $e$-folding time that is informed by a log-uniform prior. The free parameters for this model also include the stellar mass formed by the galaxy and the maximum age of the stellar population. 

\textbf{Delayed$-\tau$ + burst}: This is the same as above but including a random burst of star formation. During a burst, some fraction of mass is formed in an instantaneous burst of star formation. 

\textbf{Constant}: The constant star formation history model is set to a uniform value for all times. These are often used for modeling star forming galaxies at high redshift.

\textbf{Nonparametric}: The nonparametric SFHs as implemented in {\sc prospector} fit for the fractional stellar mass formed in a particular time bin, independent of the total mass formed (i.e. the shape of the SFH is separate from the normalization). For this model, the marginal probability distribution on the specific star formation rate in each bin follows a Dirichlet prior centered on a constant SFR(t). The time bins can be adjusted in both number and size by the user, but remain fixed during the fit. An additional parameter, called the concentration parameter, is required to specify the prior and controls the concentration of stellar mass formation across time bins. Lower concentration values ($\sim0.2$) result in more bursty star formation histories, while higher values ($\sim1.0$) result in smoother SFHs. We chose a value of $0.7$ to allow for short timescale variations in the SFH, though we briefly explore the impact of this prior choice in Appendix \ref{app:nonpara}. We also choose $10$ time bins spaced logarithmically in time motivated by \cite{ocvirk_2006} who found that the ability to distinguish separate stellar populations is proportional to their difference in age in logarithmic time. The last two bins do not follow this prescription and instead span the previous $100$ Myr and $100-300$ Myr, allowing for a minimally young population of stars. The choices for this model are motivated largely by the the impact of each choice on the priors for sSFR and stellar age: priors that are too narrow in inferred property space can result in biased estimates while priors that are too wide can hinder model convergence. Though this model depends on the choice of time bins, as shown in Appendix A of \cite{leja_how_2019} and our Appendix \ref{app:nonpara}, the stellar masses inferred from this model are robust against perturbations in the number and spacing of time bins while the assumed prior on the fractional masses has a much larger impact. We will refer to this model as 'nonparametric' throughout our following analysis. 

\subsubsection{Metallicity and Dust}\label{sec:dust}

Dust and metallicity also significantly impact the SED of a galaxy. Below we describe the models for each component. The same models and parameter prior distributions are used for each of the SFH model fits. 

\textbf{Metallicity}: Following \cite{leja_older_2019}, we use a prior on the stellar metallicities as a modified version of the stellar mass$-$stellar metallicity relationship from $z = 0$ Sloan Digital Sky Survey (SDSS) data (\citealt{gallazzi_ages_2005}). In practice, {\sc fsps}, and subsequently {\sc prospector}, will effectively assume a uniform metallicity across the galaxy for all time, constrained by the stellar mass$-$stellar metallicity relation. This is in contrast to the metallicity and chemical enrichment history present in the {\sc simba} simulation and {\sc prospector} stellar SEDs, where each star particle in a galaxy has an individual metallicity that contributes to the galaxy's overall enrichment history. This complexity has yet to be implemented in most SED fitting codes in part due to the degeneracies arising between metallicity, dust, and stellar age. 

\textbf{Dust Emission}: Constrained by energy balance, where the thermal emission from dust in the far-infrared is assumed to be equal to the stellar light absorbed by dust \citep{da_cunha_simple_2008}, dust emission is modeled following \cite{draine_infrared_2007}, which describes dust emission using three parameters: $U_{min}$ which is the the minimum radiation field strength in units of the MW value, $q_{PAH}$ which is the mass fraction of dust in PAH form, and $\gamma$ which is the fraction of dust in high radiation fields. $U_{min}$ and $\gamma$ are allowed to vary in the fits, but the PAH mass fraction is fixed to the true value (5.86\%) because the synthetic PAH spectra are not sampled. 

\textbf{Dust Attenuation}: The dust attenuation model is fixed to match the {\sc powderday} dust screen model. Thus the SEDs have a uniform dust component with an optical depth of $0.7$ affecting all stars and a birth cloud dust component affecting young stars (less than $10$ Myr old) with an optical depth of $0.7$.

\section{Results} \label{sec:results}

We fit SEDs for all SFH models described above. The following sections detail the results of the output of each SED fit from the fiducial run (i.e. with dust attenuation fixed to the true model). In general, we find that the commonly used parametric SFHs struggle to reproduce the physical properties of the {\sc simba} galaxies, including the stellar mass, mass-weighted age, and the recent rate of star formation. For instance, the three parametric models systematically under-estimate the stellar masses of the {\sc simba} galaxies by 0.4 dex on average. The properties mentioned above are directly dependent on the assumed SFH model: the stellar mass is the integral of the SFH across time (modulo the fraction of stars that now exist as stellar remnants and mass loss from post main sequence stars) and will depend on the amplitude of the SFH across time. The mass-weighted age of the galaxy will depend on the the shape of the SFH and the SFR depends only on the SFH over the last 100 Myr. We examine the efficacy of each SFH model to recover the above properties in the following sections.

\subsection{Stellar Mass Recovery}
\label{section:smass}

\begin{figure*}[!h]
  \centering
  \subfloat[]{\includegraphics[width=\textwidth]{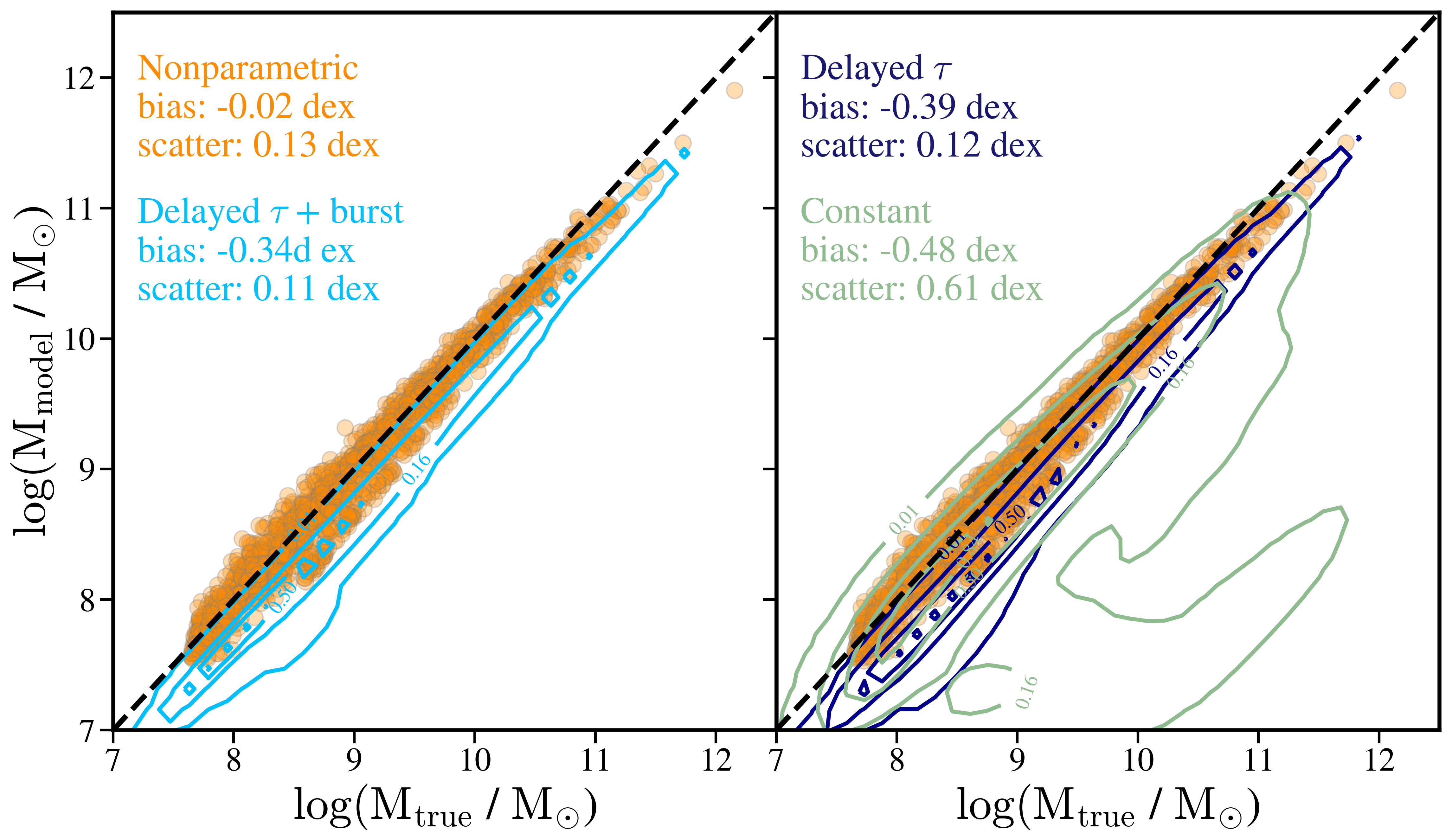}}\hfill
  \subfloat[]{\includegraphics[width=0.5\textwidth]{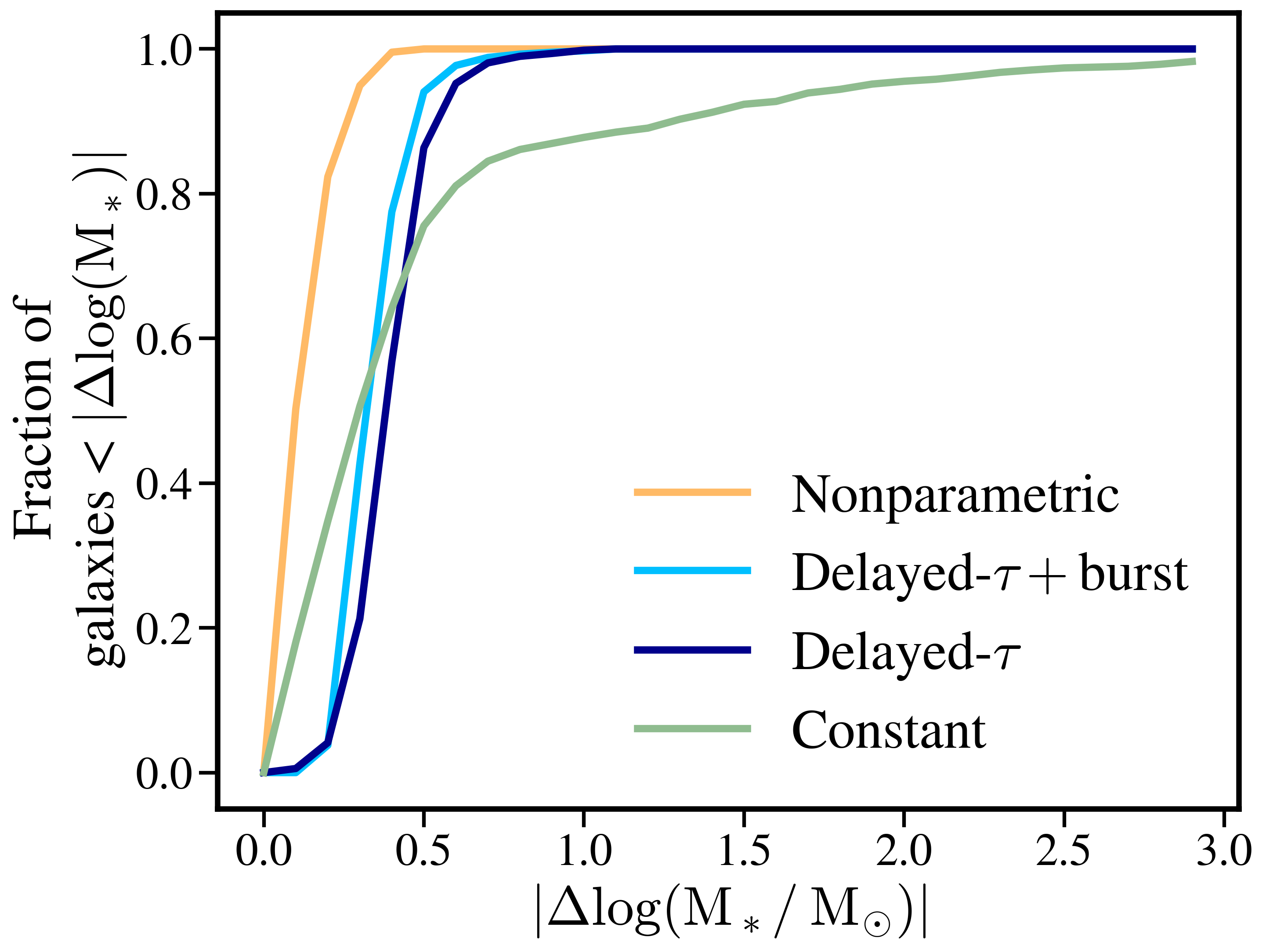}}
  \subfloat[]{\includegraphics[width=0.5\textwidth]{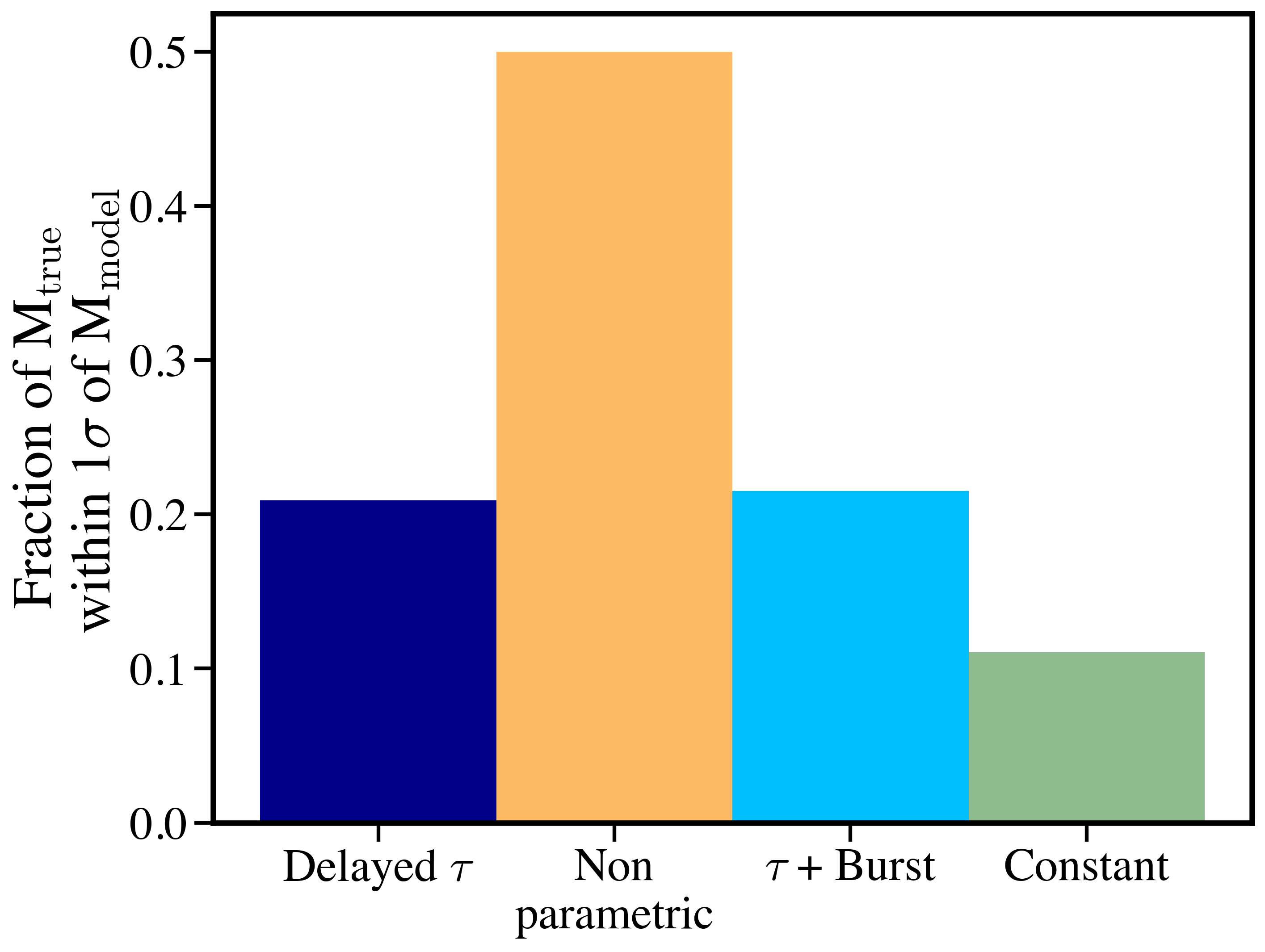}}
  \caption{\textbf{Top}: Comparison of inferred stellar M$_*$ to true stellar M$_*$ of {\sc simba} simulated galaxies for all SFH models. The bias is the average offset between the inferred stellar mass and the true. The scatter is the 1$\sigma$ standard deviation of this distribution. The masses inferred from the nonparametric SFH are shown in orange. Light blue contours show the delayed$-\tau$ + burst SFH (a parametric model). Right panel is the same as left but green contours are for the constant SFH model and dark blue contours show the delayed$-\tau$ model. Contour levels for the three parametric models highlight the $16^{th}$, $50^{th}$, and $84^{th}$ percentiles. The black dashed line is the 1:1 relation (i.e. the ideal case where the inferred mass perfectly matches the true mass). \textbf{Bottom Left}: Cumulative fraction of galaxies with inferred M$_*$ offsets. The stellar mass offsets are calculated as the absolute value of the difference between the $\log$(inferred M$_*$) and $\log$(true M$_*$). \textbf{Bottom Right}: Fraction of true galaxy M$_*$ that are within $1\sigma$ of the median inferred M$_*$ for each SFH model, where $1\sigma$ includes the $16^{th}$ through $84^{th}$ percentiles of the stellar mass posteriors. The stellar mass PDFs inferred from the nonparametric model capture the true stellar mass for more than 50\% of galaxies, compared to just 20\% for the $\tau$ models. }
\label{fig:mass_comp}
\end{figure*}

In Figure~\ref{fig:mass_comp}, we examine the impact of the star formation history model on the derived stellar mass ($M_*$) of our galaxies. In the top plot, we show a comparison between the stellar masses inferred from the various SFH models described above. The contours show the delayed-$\tau$, delayed-$\tau+$burst, and constant parametric star formation histories, while the orange points show the distribution of galaxies when using the flexible nonparametric SFH model. We compare the derived M$_*$ on the y-axis to the true M$_*$ on the x-axis. The derived quantities are the median of the stellar mass PDF for each galaxy. The stated biases for each model is the average offset between the inferred stellar mass and the true value. The scatter is the 1$\sigma$ standard deviation of this distribution. 

The nonparametric models recover the true stellar masses with significantly better accuracy afforded by the flexibility of the SFH model. To quantify this, we use two plots to show the improved accuracy and uncertainty estimates afforded by the nonparametric model. On the bottom left, we show the cumulative inferred stellar mass offset for each SFH model. The stellar masses inferred from the nonparametric model have are all within $0.4$ dex of the true stellar mass, compared to the stellar masses inferred from the $\tau$ models which suffer larger offsets. On the bottom right, we show the fraction of true stellar masses contained within the $1\sigma$ region of each galaxy stellar mass posterior, i.e. the true stellar mass falls between the estimated $16^{th}$ and $84^{th}$ mass percentiles. In other words, the uncertainty quantified by the stellar mass posterior width includes the true stellar mass. While the stellar mass PDFs inferred from the nonparametric SFH capture the true stellar mass for 50\% of galaxies, the PDFs from the $\tau$ models include the true stellar mass value for only 20\% of galaxies. The other 80\% of galaxies have a stellar mass that is offset from the true value with error bars that do not capture the true value. Thus the reported model uncertainties do not reflect the systematic biases imposed on the inferred galaxy properties. 

These three plots demonstrate the significantly improved accuracy afforded by the nonparametric SFHs as compared to the traditional parametric forms. Though M$_*$ is traditionally considered the most robust property derived from SED fitting, the results here paint a different picture: parametric SFHs fail at recovering the true stellar mass for a majority of galaxies, across all stellar mass ranges, even when we fix the dust attenuation model to be the same in the SED generation and SED fits. Furthermore, the uncertainties associated with the inferred physical properties from parametric SFHs tend to be under-reported because the model priors and subsequently posteriors are highly informative. Meaning, stellar masses inferred from parametric SFH models will be systematically under-estimated with under-reported uncertainties (i.e. the uncertainties do not increase with increasing bias). The average offset in stellar mass for the nonparametric model is $-0.02$ dex with an average uncertainty of $0.11$ dex, whereas for the delayed-$\tau$ model, these values change to $0.38$ dex and $0.19$ dex respectively. 

\subsection{Star Formation History Recovery}\label{section:sfh_rec}

\begin{figure}[h]

\centering
\includegraphics[width=0.47\textwidth]{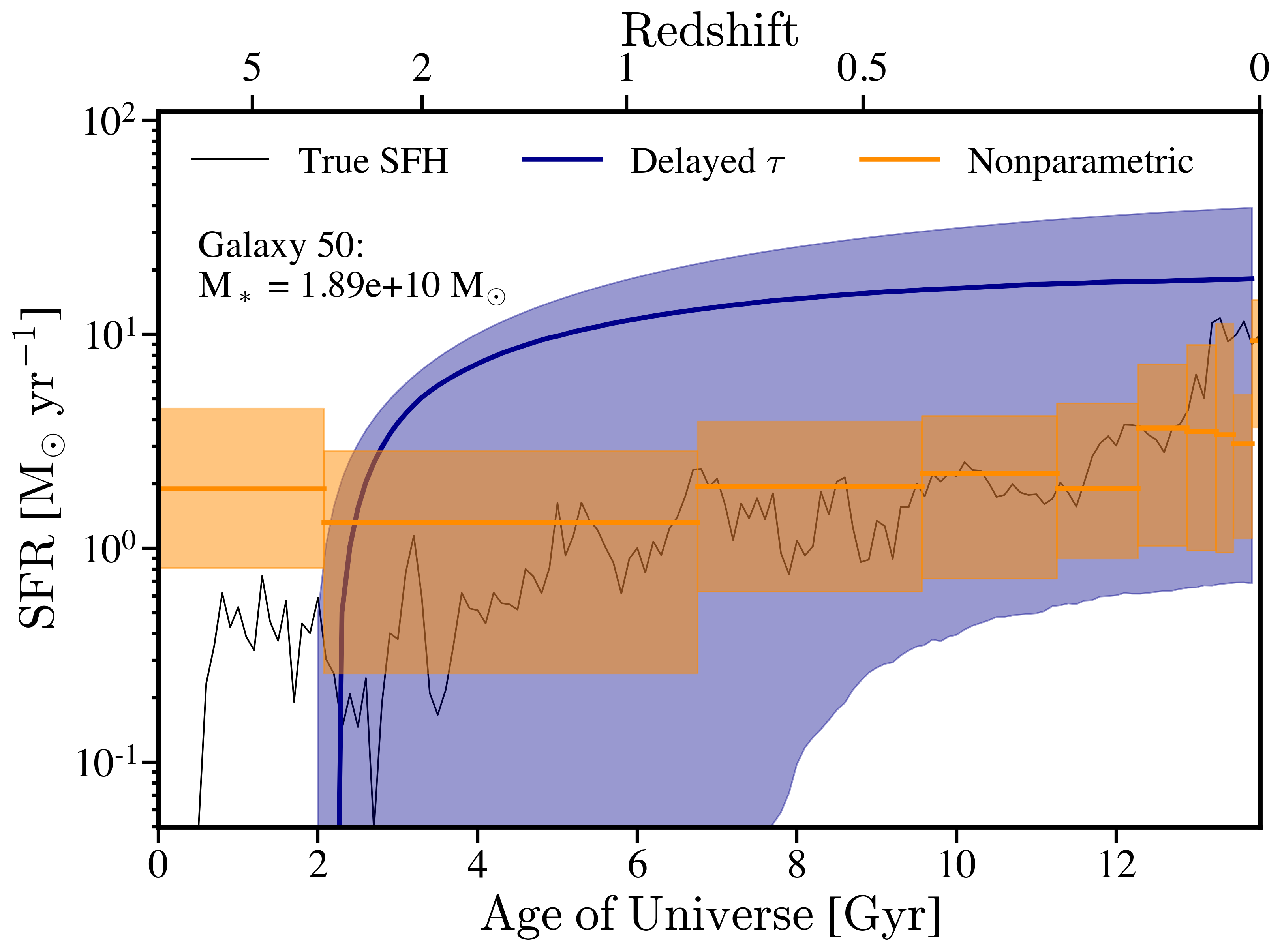}

\caption{Star formation history for an example galaxy. The true galaxy SFH is shown in black. Two of the model SFHs are shown, the nonparametric in orange and the parametric (delayed-$\tau$) is shown in blue. The $50^{\mathrm{th}}$ percentile value is shown as the solid line while the shaded regions include the $16^{\mathrm{th}}$ through $84^{\mathrm{th}}$ percentiles.}
\label{fig:sfh}

\end{figure}

A natural question is how well a given model recovers the true star formation history of a galaxy. In Figure~\ref{fig:sfh}, we show a randomly chosen galaxy's SFH and compare the recovered SFH for both a parametric and nonparametric SFH model. For this particular galaxy, the nonparametric model reasonably matches the true SFH with the median fit (solid, orange line), and the $1\sigma$ region (orange shaded region) also capturing the stochastic behavior in the true SFH over short time scales. The median delayed$-\tau$ model, however, fails to match the amplitude of the true SFH over much of cosmic time, which will result in an inferred stellar mass that is over-estimated. The $1\sigma$ region for the delayed$-\tau$ also covers the true SFH but shows the large dispersion in SFH solutions throughout the fit. 

That the nonparametric star formation histories are more accurate at recovering the stellar mass of a galaxy is mainly attributed to the fact that they are significantly more flexible and thus better at describing the various star formation histories seen in the {\sc simba} galaxy formation simulation. With only a small number of parameters describing the width and location of the curve, the three parametric SFHs (delayed-$\tau$, delayed-$\tau$ with a burst component, and constant), struggle to match the true SFH for most galaxies. This will affect not only the stellar masses inferred from each model but also the stellar ages and SFRs. The two delayed-$\tau$ models struggle to match the true SFHs that rise over time, as only very large values of $\tau$ allow for a slower decline at late times. 

For massive galaxies (M$_*$ $>$ $10^{11} \mathrm{M_{\odot}}$) at $z = 0$, the exponential decline of the parametric SFHs can match the true galaxy SFHs at late times as these massive galaxies are typically quenched or quiescent, but only at the expense of missing the large, extended early periods of star formation and thus missing out on the bulk of formed stellar mass. The lower mass galaxies tend to be bluer, star forming galaxies with SFHs ill-suited for the exponential decline at late times, so that the true SFHs are not well recovered and the stellar mass estimates will be worse, a point confirmed by Figure~\ref{fig:mass_comp}.

\begin{figure}[h]

\centering
\includegraphics[width=0.47\textwidth]{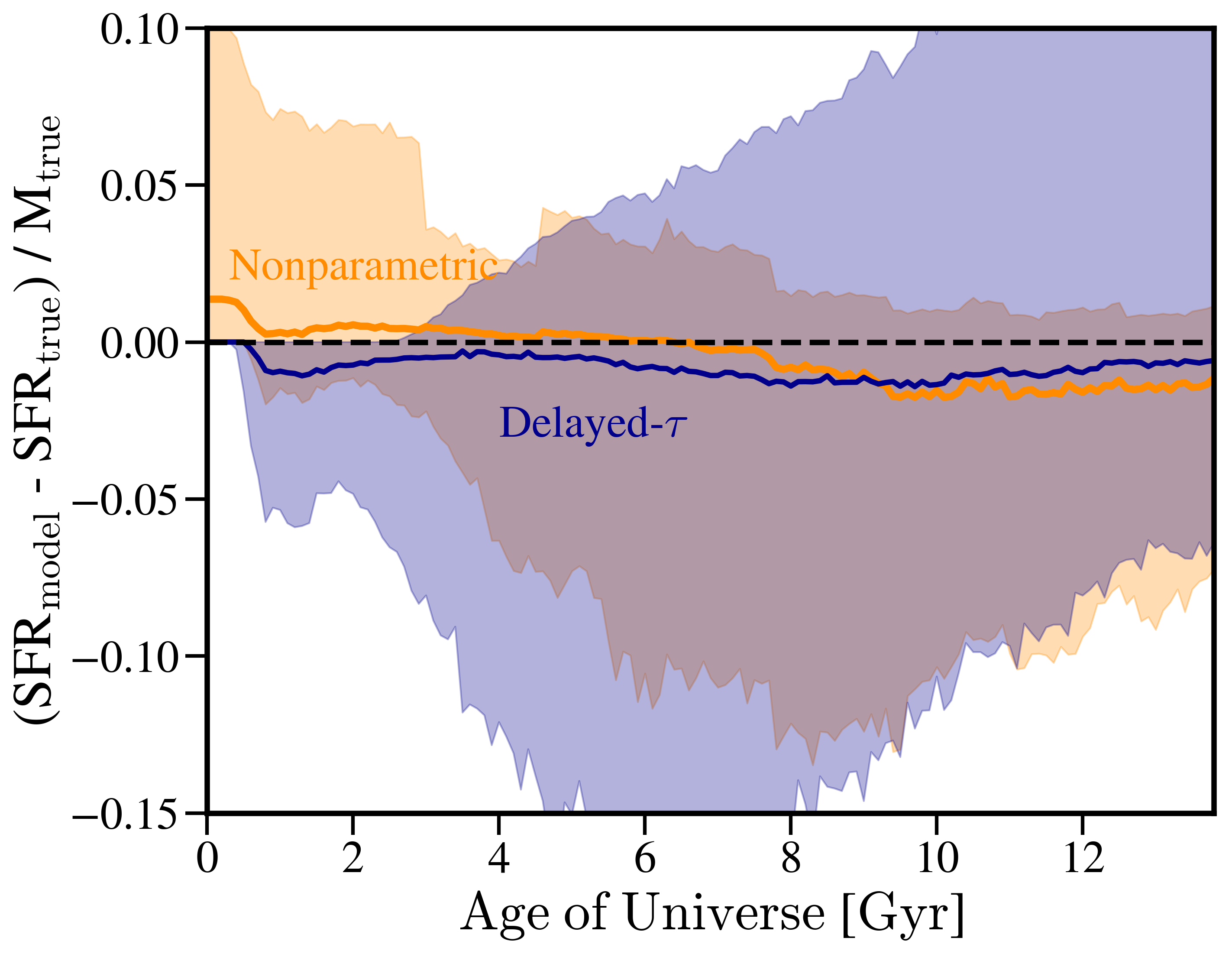}

\caption{Comparison of the SFH residuals normalized by stellar mass of all galaxies for the nonparametric model and the delayed-$\tau$ model across time. The black dashed line represents the ideal scenario of a perfect match. Offsets were calculated between the median model SFH and true SFH every 100 Myr across the entire history. The solid lines refer to the median offset for each model while the shaded regions include the $16^{\mathrm{th}}$ through $84^{\mathrm{th}}$ percentiles. The nonparametric model outperforms the delayed-$\tau$ model on average for all times.}
\label{fig:sfh_diagnostic}

\end{figure}

Building on Figure~\ref{fig:sfh}, in Figure~\ref{fig:sfh_diagnostic}, we compare the offsets between the inferred SFHs and the true SFHs for all galaxies again for the nonparametric model and the delayed-$\tau$ model. The solid lines refer to the median offset for the entire galaxy distribution for each model while the shaded regions include the $16^{\mathrm{th}}$ through $84^{\mathrm{th}}$ percentiles. The offsets were calculated between the median model SFH and the true SFH in 100 Myr intervals over the entire history. The median SFH offsets for both models are centered around zero for most of cosmic time, but the delayed-$\tau$ model has a much larger dispersion. \cite{iyer_reconstruction_2017} found similar results with the Dense Basis nonparametric SFH method for stellar mass, SFR, and stellar age when fitting mock broadband SEDs from simulations and a sample of galaxies from CANDELS and comparing these results using SpeedyMC \citep{speedymc_2011, speedymc_2015}. 

\begin{figure*}[t]
  \centering
  \subfloat[]{%
    \includegraphics[width=0.32\textwidth]{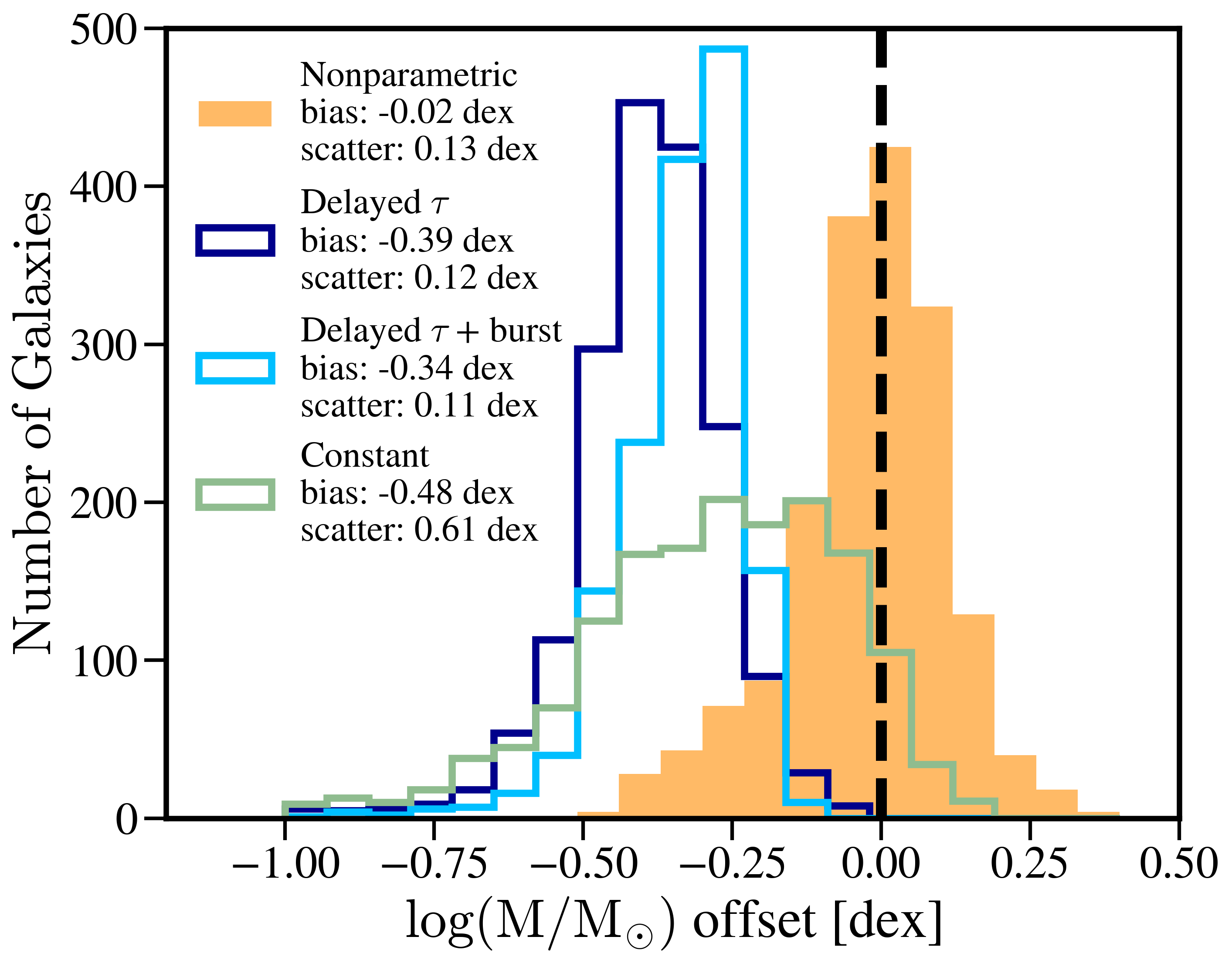}}\quad
  \subfloat[]{%
    \includegraphics[width=0.32\textwidth]{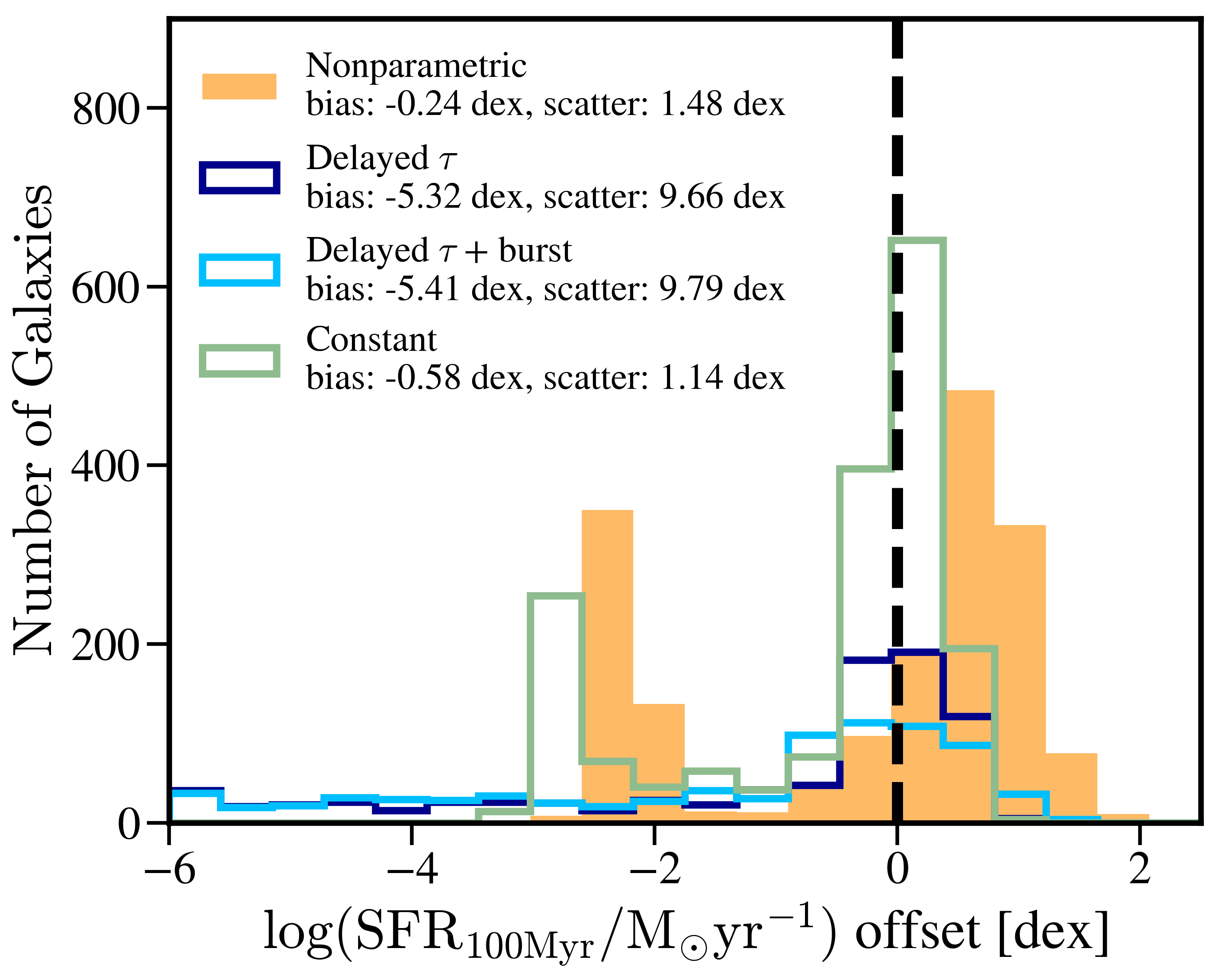}}\quad
  \subfloat[]{%
    \includegraphics[width=0.32\textwidth]{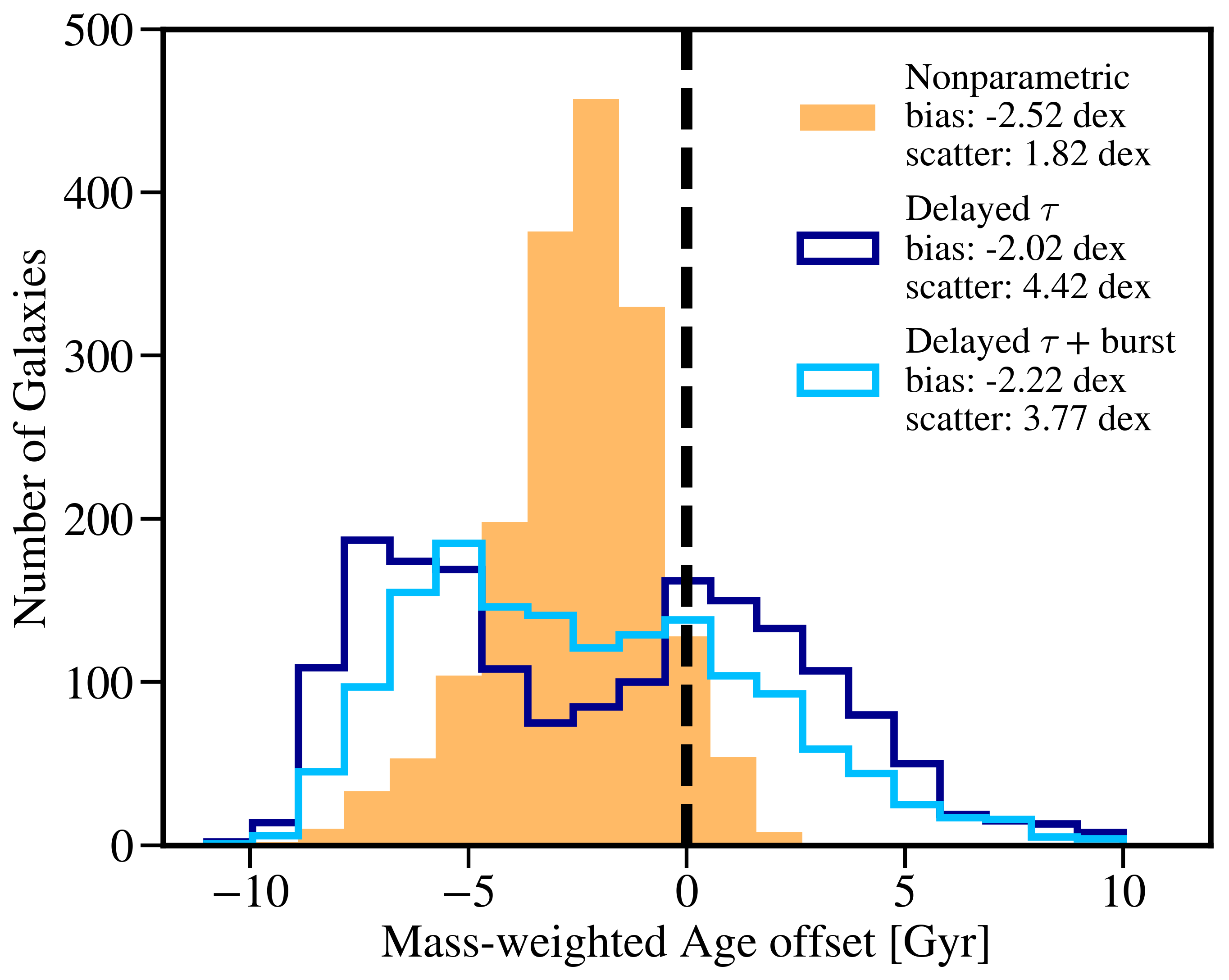}}
    \caption{Offsets from the true values of inferred galaxy properties. Properties inferred from the nonparametric model lie much closer to offsets of $0$ than the parametric models. The stated biases for each model is the average offset between the inferred galaxy property and the true value. The scatter is the 1$\sigma$ standard deviation of this distribution. \textbf{Left}: Inferred stellar mass offsets for each SFH. 13\% of galaxies fit with a constant SFR have M$_*$ offsets $>$ 1 dex. \textbf{Middle}: Same as left but for star formation rate over the last 100 Myr. \textbf{Right}: Same as left and middle but for mass-weighted age. Note that the mass-weighted age for the constant SFH is 0.5$\times t_H$, so we neglect plotting this inferred distribution.}
    \label{fig:biases}
\end{figure*}

\subsection{Ages and Star Formation Rates}\label{section:ages}

The mass-weighted stellar age of a galaxy depends on the shape of the SFH, and the accuracy these inferred properties therefore depends on the model SFH accurately matching the true galaxy SFH. We show the offsets from the true values for the mass-weighted stellar ages and star formation rates, along with stellar mass, in Figure \ref{fig:biases}. The offsets are defined as the difference between the median inferred value for each property and the corresponding true value for each galaxy. The M$_*$ offsets are shown in the left, alongside the offsets in star formation rate over the last $100$ Myr in the middle, and the mass-weighted stellar age on the right, derived from the average of the inferred SFHs over the last 100 Myr and the full SFH, respectively.

The nonparametric model shows significantly better fits for each galaxy property when compared to the parametric models. The stellar mass is the most robustly inferred property, followed by the SFR and mass-weighted stellar age. The parametric models struggle to match all three properties simultaneously. For example, the constant SFH tends to capture the late time SFR of the {\sc simba} galaxies, but the mass-weighted stellar ages cannot be accurately derived as the age inferred from a constant SFH is just 0.5$\times t_H$. All three parametric models systematically under-estimate the stellar mass and mass-weighted age of all galaxies. This bias is a consequence of the use of strong priors and the behavior of these priors when fitting photometry that tends to prefer younger stellar populations, as shown in Figure \ref{fig:sfh_priors} and \citep{carnall_how_2019, leja_how_2019}. A drawback of the delayed-$\tau$-like models is the trade off between correctly inferring stellar age or SFR for a declining SFH. Unless the prior space is manipulated to allow for a rising SFR by, e.g., allowing for very large $\tau$ values (as in the case of \citep[e.g.][]{acquaviva_2011_hightau, Ciesla_2017, aufort_2020}), the SFRs will be biased low. On the other hand, the SFRs may be correctly inferred if the peak of the SFH is placed relatively close to the time of observation, but this will occur at the expense of inferring the correct stellar age.

\section{Discussion}\label{sec:disc}

On average, the nonparametric SFH model outperforms the parametric SFHs on all metrics, including recovering the M$_*$, mass-weighted stellar ages, and late time star formation rate of the {\sc simba} galaxies. The flexible nonparametric model used here is able to more accurately infer the physical properties and growth histories of galaxies from the {\sc simba} cosmological simulation. The flexibility of the nonparametric model, compared to the relatively inflexible parametric models, is twofold: (1) the time resolution allows the SFH model to describe the shorter timescale fluctuations in the galaxy SFH and (2) the prior on the fractional SFR in each time bin results in effectively unbiased priors on sSFR and stellar age. Though the priors on these properties are not flat, they are unbiased in the sense that the center of the prior distributions do not prefer high or low values for each property; the median of the mass-weighted stellar age distribution is centered on the \cite{madau_dickinson_14} estimate of the median stellar age derived from the cosmic star formation rate density while the median sSFR is centered at $\log(1 / t_{univ} ) \sim -10.1 /$ yr which corresponds to a constant SFR. This latter point is in contrast with the $\tau$ models, where younger stellar populations are preferred due to the priors imposed on the shape of the SFH. 

Moreover, the danger in applying a $\tau$ model to a sample of galaxies lies in the false constraints on galaxy properties imposed by the SFH priors. This results in biased inferred galaxy properties with under-estimated uncertainties (i.e. the uncertainties do not increase with bias). In this section, we discuss the ways in which each SFH model impacts the inferred galaxy properties and why some models perform better than others. We also discuss the importance of carefully chosen priors when fitting SED photometry, for both nonparametric and parametric SFH models, and briefly discuss how including diffuse dust in this analysis impacts the inferred stellar masses.

\begin{figure*}
\centering
\includegraphics[width=\textwidth]{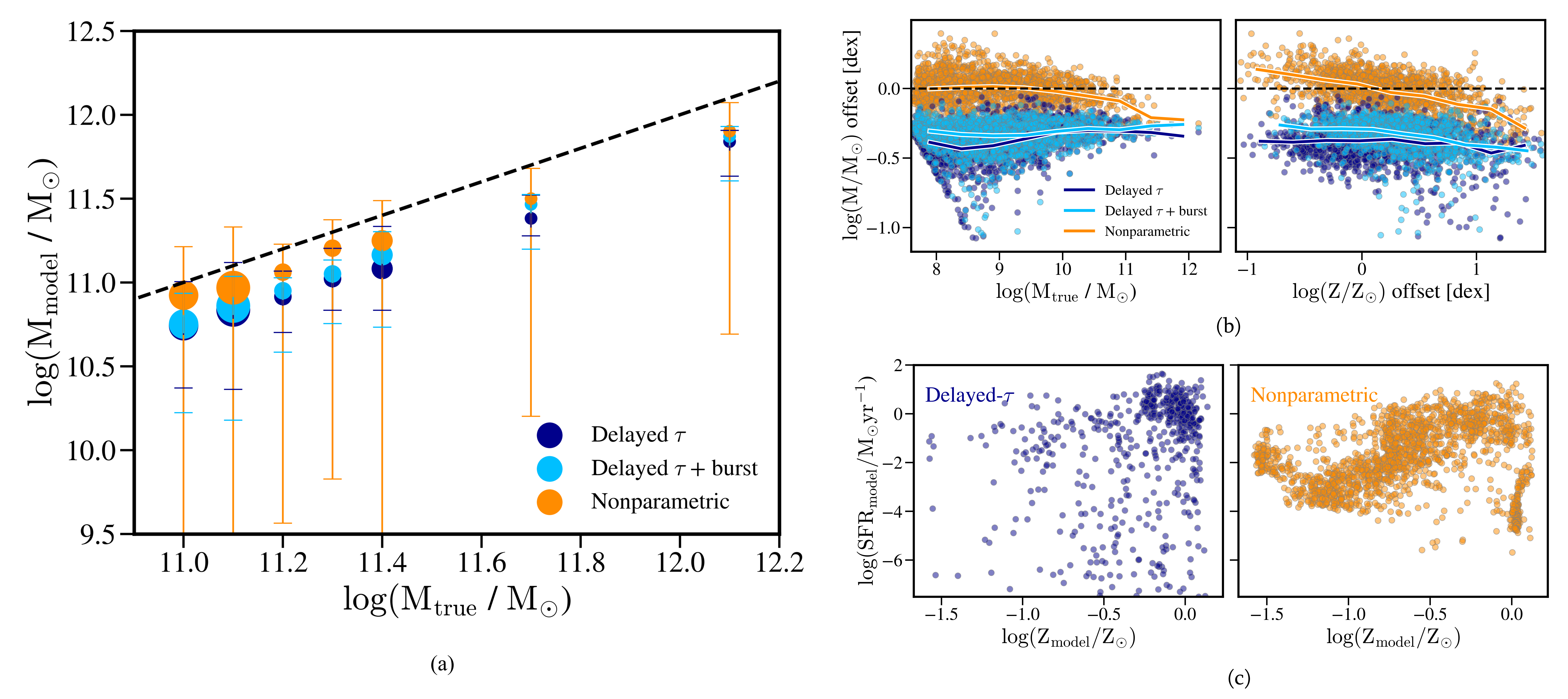} 

\caption{\textbf{left}: A binned 1:1 stellar mass comparison including uncertainties for high mass galaxies. The sizes of the points reflect the number of galaxies in each bin. The average uncertainties for each mass bin cover the biases for the nonparametric model, while the uncertainties are smaller than the biases for the parametric models for the highest mass bins. \textbf{Top Right}: Stellar mass offset as a function of true stellar mass (left) and metallicity offset (right). \textbf{Bottom Right}: Inferred SFRs as a function of inferred metallicities for the delayed-$\tau$ and nonparametric SFH models.} 
\label{fig:mass_offsets}

\end{figure*}

\subsection{Parametric Models}

All three parametric SFH models considered (constant SFR across time, delayed-$\tau$ exponentially declining SFH, and delayed-$\tau$ with an additional burst component) have been shown to systematically under-estimate the stellar masses of the {\sc simba} galaxies. This is true for galaxies of all masses and specific star formation rates. We show the stellar mass offsets as a function of galaxy stellar mass in the top right panel of Figure \ref{fig:mass_offsets} for the two delayed-$\tau$ models (the constant model is neglected for the sake of clarity). The solid lines refer to the running median of the stellar mass offset distribution for each model. The $\tau$ models systematically under-estimate the stellar masses by approximately $0.4$ dex for all galaxies. Compared to previous studies on the effect of assuming a parametric SFH model \cite[e.g.][]{pforr_2013, carnall_how_2019}, our inferred stellar mass offsets for the $\tau$ models are similar. \cite{pforr_recovering_2012} found average stellar mass offsets of $0.6$ dex for mock galaxies at z$\sim$0.5 when reddening effects were considered. Ruling out unrealistic dust and age solutions lowered these median offsets to $0.2 $ dex. Though we could, in principle, apply a correction for the offset in the $\tau$ models, we choose not to do so because such a correction would depend on the model assumptions made here, both in the SED fitting procedure and in the {\sc simba} model, and because the scatter of the tau stellar mass distributions are sufficiently large. Our aim of this analysis is not to provide correction estimates for SED models or to advocate for any one set of SED models or parameters but to provide ground-truth tests for these assumptions. Additionally, for real observations, due to the complex degeneracies in SED fitting, an inaccurately inferred SFH typically means the inferred dust, metallicity, nebular properties, etc. are also wrong. An overall correction applied to mass is not only dependent on the model assumptions in this analysis, but ultimately it is unsatisfactory because stellar mass is not the only output that depends on an accurate SFH and would otherwise negatively impact results inferred for real observations.

The inability of the parametric models to accurately recover the stellar mass of the {\sc simba} galaxies is a consequence of the restrictive nature of these models. We find that unless the inferred $e$-folding time is sufficiently large to allow prolonged star formation, especially at early times, the stellar masses recovered from these models will be under-estimated. A by-eye analysis of the library of star formation histories measured from the {\sc simba} simulation show that the $\tau$ and delayed-$\tau$ SFH models are a poor match for a majority of the star forming galaxy population at redshift $z = 0$ unless the $e$-folding time is sufficiently large ($\tau \sim$ 5 $Gyr{-1}$) to model the slow decline of quenching or to model a rising SFH. This is a consequence of the specific galaxy formation models implemented in {\sc simba} but is not unique to {\sc simba} or galaxy formation simulations in general. While a delayed-$\tau$ SFH model may match the SFR over time of an isolated closed-box galaxy (i.e. no gas inflows or outflows such that star formation is an exponential function of the in-situ gas mass and gas depletion time), it is not flexible enough to describe the shorter timescale fluctuations seen in both simulated and observed SFHs. Flexibility in the form of additional bursts of SFH or the use of a lognormal model with a large volume of parameter prior space to sample from as presented in \cite{diemer_log-normal_2017} can result in more accurate stellar masses, but do not remedy the wide dispersion in late-time SFR and stellar age as discussed below. And though the photometry inferred from each model is a reasonable fit to the observed SEDs, the inferred stellar masses differ from the true values by $-0.38$ dex on average. Again focusing on the $\tau$ models, the inferred value of $\tau$ and the placement of the peak of the SFH will impact the inferred stellar age and recent SFR. These models can also under-estimate the recent SFR of the galaxy by several orders of magnitude, evidenced by Figure \ref{fig:biases}(b), and severely under-estimate the age of the galaxy in order to match the observed SFR, evidenced by Figure \ref{fig:biases}(c). These biases are driven by the strong priors such that the parametric model is forced between including either recent star formation or an older population of stars, but not both unless the model is distorted and forced away from an exponential decline. The strongly peaked priors are also responsible for the uncertainties that do not increase in step with biases. We show in the left panel of Figure~\ref{fig:mass_offsets} a binned 1:1 stellar mass plot, zoomed in to show galaxies in the highest mass bins. The size of the points represents the number of galaxies in each bin. The uncertainties for a majority of the bins do not compensate for the large biases seen with either $\tau$ model.

\subsection{Nonparametric Model}

Again in the top right panel of Figure \ref{fig:mass_offsets}, we show the stellar mass offsets for the nonparametric SFH model in orange. This model, in contrast with the parametric models, achieves much more accurately inferred stellar masses for galaxies of all masses. The average stellar mass offset improves from $0.38$ dex with the delayed-$\tau$ model to under $0.1$ dex for the nonparametric model. We note the increasing trend the magnitude of offsets for high mass galaxies (M$_*$ $>$ $10^{11} \mathrm{M}_{\odot}$). However, only 20 galaxies populate this region of stellar mass space so it is difficult to draw conclusions about the performance of this model on high mass galaxies. We show a zoomed-in 1:1 stellar mass plot for these galaxies in the left panel of Figure~\ref{fig:mass_offsets}. Though the stellar mass offsets grow for increasing stellar masses, the uncertainties increase in step for the nonparametric model.

\begin{figure}[t]
  \centering
   \includegraphics[width=0.48\textwidth]{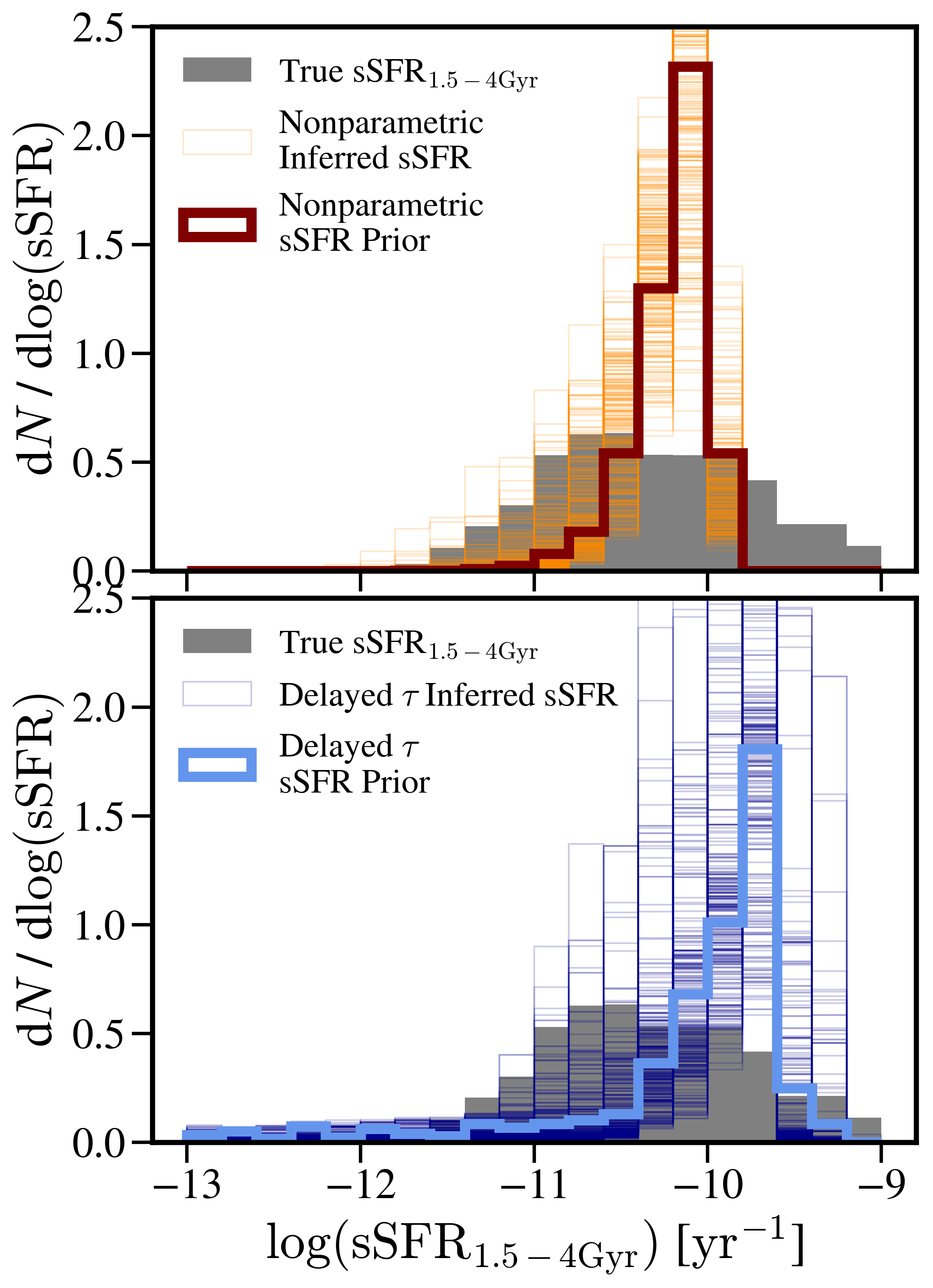}
    \caption{Distributions of the average sSFRs between the first $1.5-4$ Gyr. The true distribution for the {\sc simba} galaxies is shown by the gray histogram. \textbf{Top}: Nonparametric inferred sSFR posteriors for 300 randomly selected galaxies. The prior distribution on early sSFR is shown in the bold maroon line. \textbf{Bottom}: Same as the top panel, but for the delayed-$\tau$ model. The prior distribution is shown in the bold light blue line with individual posteriors drawn in dark blue.} 
    \label{fig:early_sfr}
\end{figure}

For the handful of high mass galaxies in this particular {\sc simba} snapshot, star formation has effectively stopped anywhere between $1$ and $4$ Gyr ago. To accurately infer the stellar mass of a galaxy, the model must match the true SFH at early times when the older stars that dominate the stellar mass are formed. However, early bursts of star formation, even for prolonged periods of time, do not leave obvious artifacts on a galaxy's SED, so accurately matching the early SFH is difficult for any SED model. Anecdotally, many of the massive galaxies from this {\sc simba} snapshot experienced prolonged periods of enhanced star formation peaking at $\sim100$ M$_{\odot}$ yr$^{-1}$ around $10-12$ Gyr ago that are not recovered well by any star formation model considered here. This is demonstrated in Figure \ref{fig:early_sfr}, where we show the distributions of the average sSFR over the first $2-4$ Gyr for the {\sc simba} galaxies compared to the inferred early sSFRs from two of the SFH models. As explored in \citep{iyer_nonparametric_2019}, inferred star formation rates from more than a couple Gyr ago are dominated by the prior set on the SFH rather than the fit to photometry, as little evidence exists in the observed SEDs of these early star forming episodes. This problem is not unique to {\sc prospector} or the nonparametric SFH model used here but to all models as SFHs are only minimally informed by broadband photometry due to lack of SED features left by old stellar populations. The distribution of early sSFRs inferred from nonparametric SFH model is narrowly peaked at log(sSFR)$\sim-10.5$, so that galaxies with smaller or larger sSFRs at early times will have inferred stellar masses that deviate from the true value as long lived stars with solar masses or lower will dominate the stellar mass content of a galaxy.

For galaxies that are actively star forming, the problem of 'outshining' will augment the above difficulties, as the massive stars that are formed recently will outshine the older stellar populations that dominate a galaxy's stellar mass such that the inferred stellar mass will be heavily dependent on the priors informing the stellar age distribution \citep[e.g.][]{papovich_stellar_2001, pforr_recovering_2012}. In other words, the stellar masses inferred for these systems are dominated by the constraints from the prior (shown in Figure \ref{fig:sfh_priors} and Figure \ref{fig:early_sfr}) in addition to the constraints from photometry. In both cases, for quiescent and actively star forming galaxies, the underlying cause of the differences in model and true stellar masses is the inability to capture early, prolonged star formation activity with the model SFH.

Besides the assumed SFH, assumptions about a galaxy's stellar metallicity and the mapping from metallicity to the SED will also impact the inferred stellar mass. In this analysis, we used the \cite{gallazzi_ages_2005} stellar mass$-$ stellar metallicity relation. However, as noted in \S\ref{sec:sed_fitting}, this relatively simplistic model does not entirely capture the growth history of the {\sc simba} galaxies as the simulated star particles have metallicities that are neither uniform nor static through time. Though we isolated our results from the inclusion of a diffuse dust component, in truth our assumptions in this metallicity model will also impact the stellar masses inferred by {\sc prospector}. In the top right panel of Figure~\ref{fig:mass_offsets}, we show the stellar mass offsets as a function of the stellar metallicity offsets. The solid lines refer to the running median of each distribution. For the sake of comparison, we take the true galaxy metallicity to be the mass-weighted stellar metallicity for each galaxy. This way, we have an aggregate metallicity to compare to the inferred stellar metallicities. We see a strong inverse correlation between the offset distribution for galaxies fit with the nonparametric model but not for either parametric model. This trend is also seen in fits where the stellar metallicity is not tied to the stellar mass. One explanation for this is the inability of the uniform and static metallicity model to recover the distribution in true stellar metallicities. A model describing the chemical enrichment history of galaxies could be implemented but would be minimally constrained by broadband photometry and most likely highly degenerate with the SFH and dust model. In the absence of dust, stars with lower inferred metallicity will have excess UV compared to stars with higher inferred metallicity. This change in UV light in the SED can impact both the inferred star formation rate, and to a lesser degree, the inferred stellar mass. In Figure \ref{fig:mass_offsets}(c) we see that a trend also exists between the inferred SFR and metallicity from the nonparametric model, where generally galaxies with larger inferred SFRs will have large metallicites. These recent SFR bins are covariant with metallicity and allowing freedom in these, which is not allowed by the tau models, introduces a covariance with stellar mass could explain the trend we see in Figure \ref{fig:mass_offsets}(b). 

\subsection{Model Priors, Uncertainties, and Degeneracies}\label{sec:degen}

An underlying theme of this analysis is the flexibility afforded by the nonparametric SFH model. This flexibility is in the form of both model variability and minimally informative (i.e. only moderately peaked) priors. Here, we briefly comment on the dependence of our results on the choice of hyper-parameters for the nonparametric model (concentration of mass formation, number of times bins). We also comment on the change in performance in the presence of additional uncertainties from the stellar population synthesis models and observational noise.

\subsubsection{Tuning of the Nonparametric Model}

The nonparametric model hyper-parameters and other priors described in \S \ref{sec:sfh} were chosen for maximum flexibility. While the choice of prior for the degree of mass concentration (i.e. large variations over short timescales vs. a smoother SFH) matters for the nonparametric SFH (Appendix \ref{app:nonpara}), we find that the results presented here are not heavily dependent on the choice of time bins. The priors on either end of the spectrum in terms of favoring smooth SFHs result in stellar mass offsets that are smaller than the parametric models but are much larger compared to the fiducial nonparametric SFH model used in this analysis. The choice of prior is equivalent to constraining the shape of the SFH; thus the strongest prior on the inferred stellar mass is set by the choice of a SFH form that \textit{a priori} assumes a certain shape, as is the case for the parametric models. Moreover, the priors on the $\tau$ model parameters (i.e. $\tau$ and the time at which the SFH peaks) generally cannot be tuned to a given data set because (i) that could exclude large regions of parameter space for the inferred properties and (ii) the parameters couple differently in different contexts. For instance, if we want to do well in recovering SFHs for galaxies with rising SFRs at late times, we could impose a prior on $\tau$ that would favor large $e$-folding times. But this would necessarily bias the results against galaxies that do not have rising SFHs. Furthermore, the influence of the priors of the parametric model parameters like $\tau$ and $t_{age}$ on the subsequent galaxy physical properties are not straightforward; setting an uninformative prior on $\tau$ will not result in a flat prior on SFR, sSFR, or stellar age. The effective priors on galaxy properties are primarily driven by the choice of a declining exponential form in the first place. Comparing the delayed-$\tau$ model to the simpler $\tau$ model, which does not allow for rising SFHs even with large $\tau$, \cite{wuyts_2011} found little difference between ages and stellar mass functions inferred from either model, especially for galaxies subjected to outshining. In other words, modifying the $\tau$ model to allow for rising SFRs did not result in significantly better results for stellar mass or age. Because the nonparametric model does not favor any SFH shape beforehand, the model can tackle the wide diversity of SFHs as seen in {\sc simba} and elsewhere.

\subsubsection{Uncertainties from Stellar Evolution and Observational Noise}

A major source of uncertainty in SED modeling originates from stellar evolution; the choice of stellar isochrones, spectral libraries, and initial mass function (IMF) contribute to large uncertainties in stellar mass \citep[e.g.][]{conroy_modeling_2013, pforr_2013, 2015ApJ...801...97S}. Our results have been independent of these uncertainties since we mirror the stellar population synthesis (SPS) models between the synthetic {\sc powderday} SEDs and the SED modeling with {\sc propsector}. One such aspect of SPS modeling is the assumed stellar IMF. Re-modeling the {\sc powderday} SEDs with a mismatched IMF (\cite{chab_imf} vs. \cite{kroupa_initial_2002}), we find that even in the presence of SPS uncertainty, the distribution of stellar mass offsets between the nonparametric model and the delayed-$\tau$ model remain clear (Appendix \ref{app:imf_noise}). However, the assumed stellar IMF is just one component of an SPS model. The assumed stellar spectral library, and whether to use an empirically based or theoretical library, along with the choice of stellar isochrones have been shown to impart serious uncertainties on inferred stellar masses \citep{conroy_2009_sps_unc}. These uncertainties stem from difficulty in sampling rare stars (e.g. massive, low metallicity) and stars in relatively short-lived phases (e.g. thermally-pulsating AGB stars) as well as difficulties in modeling stellar atmospheres to produce theoretical spectra. The issue of how to model thermally-pulsating AGB stars is of particular interest as it has been shown to largely impact inferred stellar masses since these stars dominate the near-IR luminosity of intermediate-aged galaxies \citep{maraston_2006_tpagb, conroy_2009_sps_unc}.

Further observational uncertainties on inferred stellar masses originates from photometric noise, especially for data sets containing spectral line and continuum information. The additional information provided by spectra helps to discern different stellar populations and provides more information about the metal content in galaxy, potentially helping to avoid degeneracies between these parameters. However, spectra tends to suffer from calibration issues and the information provided by spectra is dependent on the signal-to-noise of said spectra. Though we do not consider spectra from nebular emission in this analysis, we conduct a brief test of the performance of the SFH models in the presence of noise by perturbing the input broadband photometry points within the 3\% uncertainties. We show in Appendix \ref{fig:appen_imf_noise} that fits to these noisy SEDs produce results similar to the above analysis, thus the results presented here, isolated from additional sources of uncertainty, hold. However, in future work we will conduct a more in-depth investigation into how these results change as a function of varying signal-to-noise and SED coverage; and for a thorough investigation into the efficacy of these models for varying signal-to-noise ratios of mock empirical spectra, see \cite{leja_how_2019}.

\subsubsection{Uncertainties from Model Degeneracies}

How an SED model handles degeneracies between model parameters and their mapping to the observed SED is a large difference between the parametric and nonparametric SFH models studied here. For the parametric SFHs, placing a strong prior on, e.g., the shape of SFH will affect the properties derived from the model and the correlations between properties by ruling out certain parameter combinations from the beginning. Though this can break degeneracies between, e.g., dust and stellar age, it is done so through strongly peaked priors with little physical basis and results in strongly biased predictions. For instance, we see in Figures \ref{fig:sfh_priors} and \ref{fig:biases}(c) that the parametric models all favor younger stellar populations than the true ages even in the absence of realistic dust, an aspect of the $\tau$ model priors noted by \cite{carnall_how_2019}. Because many degenerate, though physically plausible, solutions are ruled out \textit{a priori} by the model choice (e.g. two stellar populations produced by two separate star bursts), the inferred uncertainties on galaxy properties like stellar mass are under-estimated, as shown in Figures \ref{fig:mass_comp}(c) and \ref{fig:mass_offsets}(a). The nonparametric SFH, on the other hand, is subject to the same degeneracies but do not impose such strong biases on inferred properties as a result of the carefully applied priors. In the Bayesian framework, an overly complex model will provide unconstrained results; thus if the chosen number of time bins is too great relative to the input information or the model solutions are highly degenerate, the galaxy properties inferred with a flexible nonparametric SFH model will return increasingly larger posteriors. We see evidence of this in the middle panel of Figure \ref{fig:mass_offsets}, where average uncertainties for some of the high stellar mass bins extend more than $2$ dex.

Focusing on the stellar mass inferred from the {\sc prospector} SED fits, the non-uniform sensitivity to variations in the SFH of a galaxy make it difficult to untangle the contribution of old stars, young stars, and metallicity, and dust to the integrated SED of a galaxy and can result in inaccurately inferred galaxy properties. It is worth restating that the results from the specific implementation of nonparametric SFHs presented here outmatch the accuracy of the three commonly used parametric forms considered, as shown in Figures \ref{fig:mass_comp} and \ref{fig:biases}. The use of more simple parametric SFH forms to determine the physical properties of galaxies will result in systematically biased stellar mass estimates with SFRs and stellar ages that are not well constrained, even in a best case scenario of $25$ broadband photometry points and a constrained dust model.

\subsection{The Impact of Diffuse Dust}\label{sec:diffuse_dust}

The results discussed so far do not consider the impact of diffuse galactic dust on the inferred stellar mass from SED modeling. Instead, we forced the {\sc powderday} dust radiative transfer simulations to employ a dust screen model around stars in order to force an apples-to-apples comparison with the SED fitting techniques that also model dust via a screen model. This allowed us to isolate the differences between the input {\sc powderday} SEDs and the output {\sc prospector} SEDs and focus on how the assumed SFH model affects the inferred stellar mass. However, by neglecting the true dust distribution and dust-to-star geometry, we under-estimate the uncertainty with which we can infer the stellar mass of a galaxy. In this section, we now briefly explore how including diffuse dust impacts the results presented so far (noting that a full exploration of the impact of dust attenuation will be presented in future work). 

\subsubsection{Diffuse Dust Radiative Transfer}
For the exploration of diffuse dust on our results, we use {\sc powderday} to perform radiative transfer on the {\sc simba} galaxies, as described in \S \ref{sec:pd}.  In this situation, however, we abandon our initial assumption of a uniform dust optical depth for all stars. Instead, the {\sc fsps} stellar SEDs are propagated through the dusty ISM. The diffuse dust content is derived from the on-the-fly self-consistent model in {\sc simba} \citep{li_dust--gas_2019}, and this dust 
 is assumed to have extinction properties following the carbonaceous and silicate mix of \cite{draine_infrared_2007}, that follows the \cite{weingartner_draine} size distribution and the \cite{draine_03} renormalization relative to hydrogen. We assume $R_{\rm V} \equiv A_{\rm V}/E(B-V) = 3.15$. We do not assume further extinction from sub-resolution birth clouds. PAHs are included following the \citet{robitaille_pahs} model in which PAHs are assumed to occupy a constant fraction of the dust mass (here, modeled as grains with size $a<20 \AA$) and occupying $5.86\%$ of the dust mass. The dust emissivities follow the \citet{draine_infrared_2007} model, though are parameterized in terms of the mean intensity absorbed by grains, rather than the average interstellar radiation field as in the original Draine \& Li model.  

\subsubsection{Impact of Diffuse Dust on the Inferred Stellar Mass}

\begin{figure}
    \centering
    \includegraphics[width=0.47\textwidth]{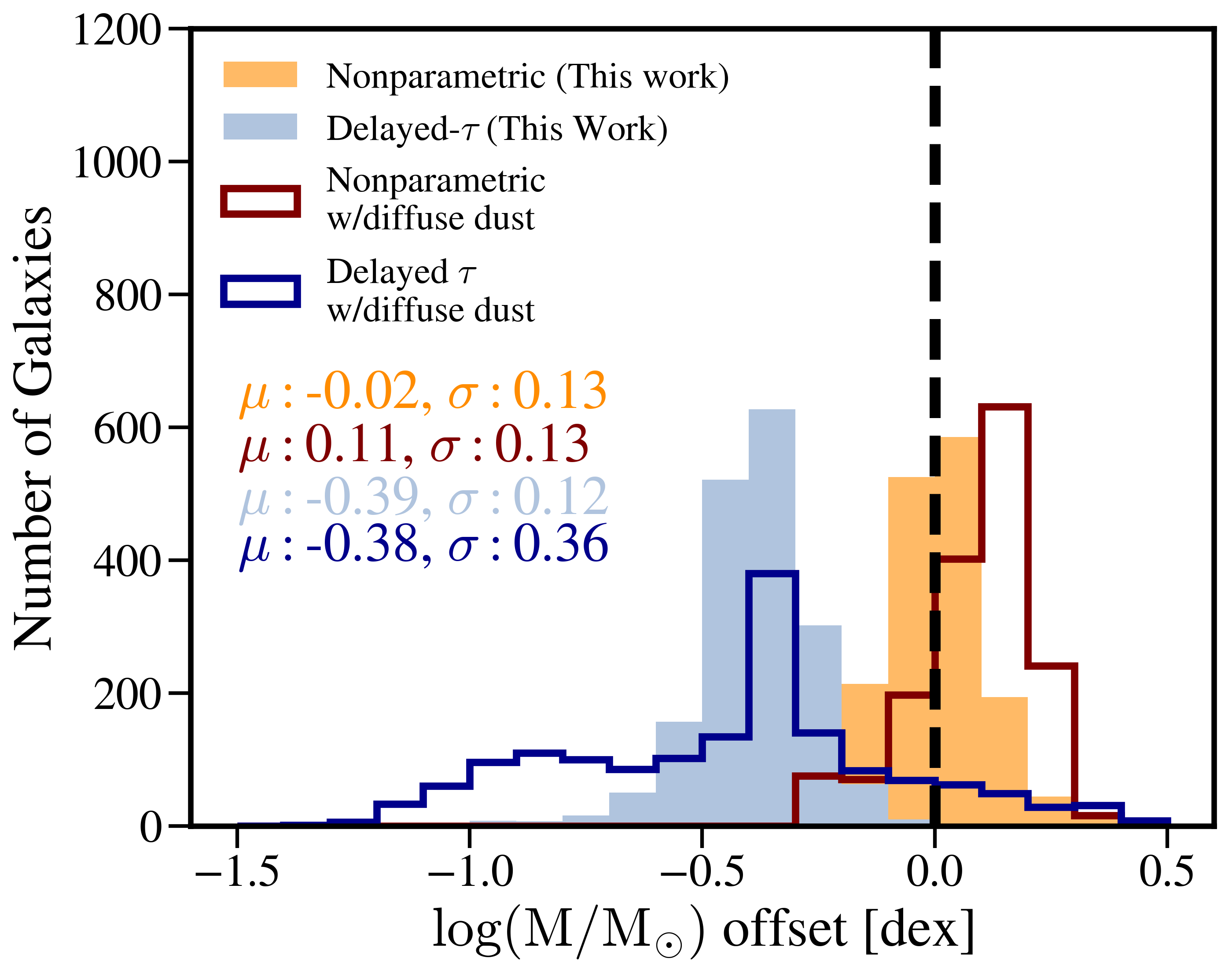}
    \caption{Distribution of stellar mass offsets for the dust screen models (filled histograms) and the models including diffuse dust and a variable dust attenuation curve (unfilled histograms). The median and 1-$\sigma$ width of each distribution is shown.} 
    \label{fig:diffuse_dust_masses}
\end{figure}

To include the impact of diffuse dust in this analysis, we again fit the {\sc simba} broadband SEDs with {\sc prospector}, this time allowing a flexible dust attenuation curve following the parametrization presented in \cite{kriek_dust_2013}. In Figure \ref{fig:diffuse_dust_masses}, we compare the stellar mass offsets from two of the SFH models with and without diffuse dust for a subset of {\sc simba} galaxies. The stellar masses inferred from the nonparametric SFH model are only marginally affected by the addition of diffuse dust. The difference in the median of the distribution of stellar mass offsets between the two data sets is statistically insignificant. The delayed-$\tau$ model, on the other hand, is significantly affected by the addition of diffuse dust: the magnitude of stellar mass offsets extends to just over $1$ dex. And while, like the nonparametric model, the median of the stellar mass offset distribution remains largely unchanged for the dealyed-$\tau$ model, the dispersion of the distribution has increased, such that a larger fraction of galaxies have larger offsets compared to the SED fits without diffuse dust. 

The addition of diffuse dust and variable attenuation curve increases not only increase the number of model parameters {\sc prospector} must fit simultaneously, but introduces degeneracies between model parameters that are difficult to untangle, even with the constraints on, e.g., stellar metallicity in place. The addition of diffuse dust amplifies the drawbacks of using a parametric SFH model shown previously. In future work, we will explore these drawbacks in detail by considering a flexible dust attenuation curve to capture the diversity of attenuation curves present in the Universe and to follow up this analysis by determining the impact of observational sources of uncertainty on inferred galaxy properties.

\section{Conclusions}\label{sec:conclusion}

We have used simulated galaxies from the {\sc simba} cosmological simulation to understand our current ability to infer the stellar mass of a galaxy with SED modeling. In particular, we assessed the impact of the star formation history model in SED fitting. We considered four SFH models, three commonly used parametric models and one nonparametric model, included in the {\sc prospector} modeling framework. We have demonstrated that the biases in stellar mass estimates decreases significantly with the use of nonparametric SFHs, falling below $0.09$ dex for a majority of modeled {\sc simba} galaxies. The conclusions from our analysis are:

\begin{itemize}
    \item Stellar masses derived from the {\sc prospector} nonparametric SFH models are much more accurate on average than those derived from parametric models, for galaxies of all stellar masses, ages, and morphologies. Figure \ref{fig:mass_comp}(a) shows the median derived M$_*$ for galaxies assuming different SFHs models. We find that the offset between the inferred M$_*$ and the true M$_*$ decrease from $0.4$ dex on average when modeled with a delayed-$\tau$ SFH to $-0.02$ dex when modeled with a nonparametric SFH. Outliers exist for galaxies at the high mass end, caused by failures to recover early periods of intense star formation. An important note is that while our results improve when using this particular nonparametric model, the differences between nonparametric and parametric begin to smear out once diffuse dust, noise, and other model uncertainties are considered. That said, we present here an estimate for the baseline M$_*$ uncertainties achievable with current broadband SED models. 
    
    \item Parametric SFHs suffer from biases that are larger than their associated uncertainties, as explored in this analysis and in, e.g., \cite{simha_parametrising_2014}; \cite{salmon_relation_2015}; \cite{carnall_how_2019} and \cite{curtis-lake_modelling_2020}. The danger in applying a delayed-$\tau$ SFH to a sample of galaxies, aside from the relatively poor match to true SFHs on average, lies in the false constraints imposed on galaxy properties by the SFH priors. As shown in Figure \ref{fig:mass_comp}(c), we find that the nonparametric SFHs tend to capture the true mass value within the 1$\sigma$ M$_*$ posterior width for a much higher fraction of galaxies compared to the three parametric models considered.
    
    \item Nonparametric SFHs in {\sc prospector} are also able to match the true {\sc simba} SFHs across time significantly better than the parametric models, as shown in Figure \ref{fig:sfh_diagnostic}. This increase in accuracy is owed to the well-constrained (i.e. through the choice of prior) flexibility permitted by the nonparametric model, so that the SFR at any one epoch is not determined by the SFR at another time. Mass recovery can be further improved by using more discerning data such as spectra or narrow band photometry, which could provide better constraints for early star formation activity.
    
\end{itemize}

Though the nonparametric SFH model used here outperformed the other parametric models on all metrics, it is important to note that these models must be used carefully. As described in \citet{leja_how_2019}, the priors that are chosen to constrain a nonparametric SFH are the primary drivers of the size of the inferred M$_*$ posterior. For the nonparametric models in {\sc prospector}, this means that the stellar mass posteriors, while wider than those of the parametric SFH models, are able to capture the true stellar mass of the simulated galaxies at much higher fractions. The {\sc prospector} nonparametric priors, including the Dirichlet prior chosen for this analysis, perform much better than the parametric SFHs across the board but the degree to which this performance improves is dependent on the choice of prior, as shown in Appendix \ref{app:nonpara}. 

The difficulty in SED modeling lies in the fact that the star formation history is only moderately constrained by broadband photometry, so priors must be carefully implemented to allow a diverse range of SFHs to be modeled while simultaneously fitting for dust and other model parameters. On this point, the {\sc prospector} nonparametric models are significantly better than the parametric SFHs used here and some more simplistic implementations of nonparametric SFHs at producing meaningful error bars thanks to the carefully chosen priors and SED modeling can only benefit from other thoughtful implementations of model priors. Finally, cosmological simulations can play an important role in future work to constrain priors not only for SFHs but also dust attenuation laws. We can also develop nonparametric models for dust attenuation in a similar way but the increase in computational resources and model degeneracies warrant caution. As such, we hope to explore further improvements to SED modeling and deriving physical properties from broadband photometry.

\section*{Acknowledgements} D.N. acknowledges support from the US NSF via grant AST-1715206 and AST-1909153, and from the Space Telescope Science Institute via grant HST-AR-15043.001-A.  J.L. is supported by an NSF Astronomy and Astrophysics Postdoctoral Fellowship under award AST-1701487. C.C. acknowledges support from the Packard Foundation.

The {\sc simba} simulation used in this work was run on the ARCHER U.K. National Supercomputing Service, {\tt http://www.archer.ac.uk}.

\textit{Software}: {\sc simba} \citep{dave_simba:_2019}, {\sc caesar} \citep{2014ascl.soft11001T}, {\sc powderday} (\citealt{narayanan15a}), {\sc hyperion} \citep{robitaille_hyperion:_2011}, {\sc fsps} \citep{conroy_propagation_2009}, {\sc python-fsps} \citep{dan_foreman_mackey_2014_12157}, {\sc prospector} \citep{2017zndo...1116491J}, {\sc numpy} \citep{5725236}, {\sc matplotlib} \citep{2018zndo...1482098C}, {\sc astropy} \citep{2020SciPy-NMeth}

\appendix
\section{Nonparametric SFH Hyper-parameters}\label{app:nonpara}

\begin{figure}[h]
    \centering
    
    \includegraphics[width=0.48\textwidth]{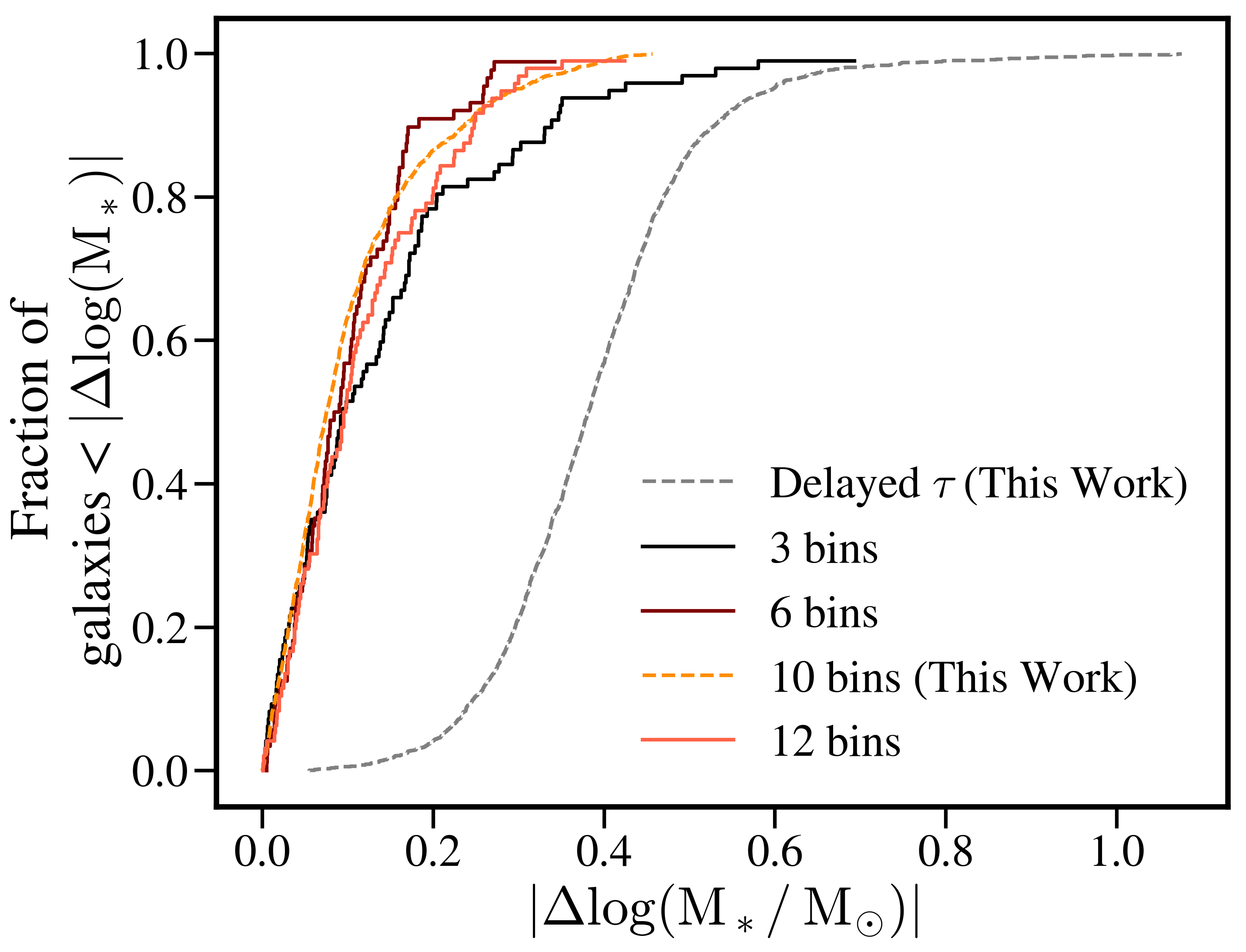}
    \includegraphics[width=0.48\textwidth]{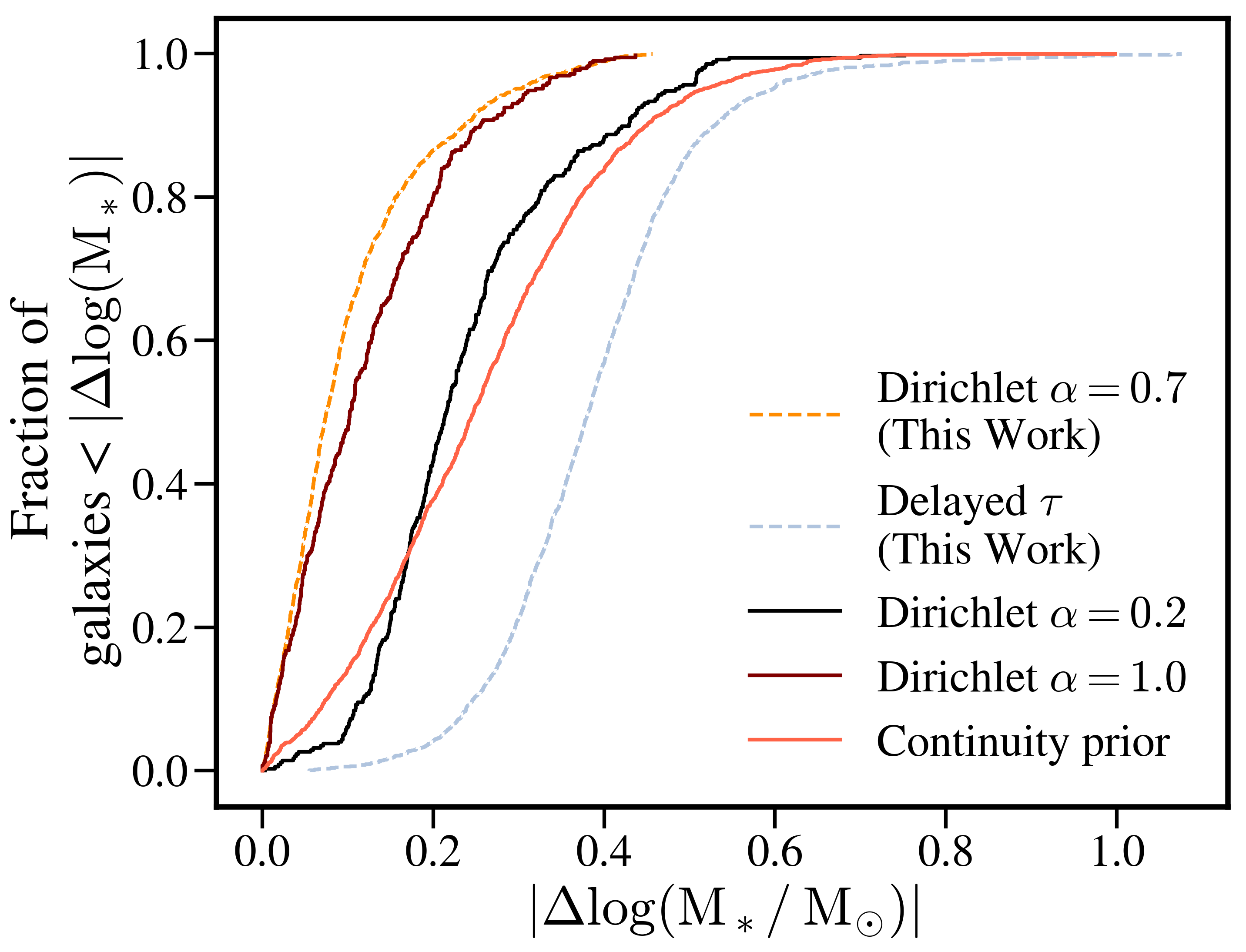}
    \caption{Cumulative fraction of galaxies with inferred stellar mass offsets for \textbf{Left}: varying numbers of time bins and \textbf{Right}: prior on fractional mass per time bin used in the nonparametric SFH model. The data presented in this analysis is shown by the dashed lines.}
    \label{fig:appen_nonpara_hp}
\end{figure}

The nonparametric SFH used in this analysis (described in full in \cite{leja_how_2019} along with all other models available in {\sc prospector}) is described by $N$ bins of constant SFR, where $N=10$ and the fraction of mass formed in each bin of star formation is constrained by a transformation of the Dirichlet prior \citep{dirich_transform} parametrized by $\alpha$, which sets the concentration of mass formation. Modulating the value of $\alpha$ modulates the preference for all stellar mass to be formed in one bin ($\alpha$ $<1$) vs. a smoother distribution of stellar mass ($\alpha$ $>1$). In practice, the fractional mass formed in each bin is fit, as opposed to the actual value of mass formed in each bin, so as to avoid sampling from a large volume of prior space and to separate fitting the shape of the SFH (done by the fractional mass) from the surviving stellar mass (the normalization of the SFH). Many tests were run to determine the dependence of the inferred stellar masses on these hyper-parameters. Figure \ref{fig:appen_nonpara_hp} shows the inferred stellar mass offsets for different choices of time bins and priors. We also include summary statistics in Table \ref{table:appen_table}. The choice of priors includes changing $\alpha$ in the Dirichlet prior along with a different prior distribution, called the Continuity prior, which favors small changes between adjacent time bins of star formation (similar to Dirichlet priors with large $\alpha$ values though star formation between bins is explicitly tied to each other). The dominant factor is the choice of prior constraining the fractional mass in each time bin, highlighting the importance of choosing a prior suited to a particular data set. 

The continuity prior was tuned in \cite{leja_how_2019} to galaxy SFHs from the Illustris hydrodynamical simulation. As recently presented in \cite{iyer_2020}, galaxy SFHs in Illustris have less power on short timescales compared to galaxies from {\sc simba}. This means that {\sc simba} galaxies tend to have more fluctuations on short timescales, so implementing a SFH prior favoring small fluctuations would tend to give worse stellar mass estimates. Furthermore, because the 'correct' choice of prior for any sample of galaxies, both simulated and observed, will change depending on the class of galaxy (star forming vs. quiescent, local vs. high redshift), we do not suggest that the particular priors we imposed on the {\sc simba} data set will be correct for all galaxies.

% \usepackage{multirow}
\begin{table}[]
\centering
\begin{tabular}{llll|lc}
SFH Model                                            &  & \multicolumn{2}{c}{Dataset / Model}                                           &  & Median M* Offset \\ \hline
\multicolumn{1}{l|}{\multirow{15}{*}{Nonparametric}} &  & \textit{Concentration Parameter}                & 0.2                         &  & 0.22 $\pm$0.12   \\
\multicolumn{1}{l|}{}                                &  &                                                 & \textbf{0.7}\footnote{Fiducial model choices are highlighted in bold.}                &  & -0.02$\pm$0.13   \\
\multicolumn{1}{l|}{}                                &  &                                                 & 1.0                         &  & 0.08$\pm$0.13    \\
\multicolumn{1}{l|}{}                                &  &                                                 & continuity prior            &  & 0.24$\pm$0.15    \\
\multicolumn{1}{l|}{}                                &  &                                                 &                             &  &                  \\
\multicolumn{1}{l|}{}                                &  & \textit{Time Resolution}                        & 3 bins                      &  & -0.06$\pm$0.21   \\
\multicolumn{1}{l|}{}                                &  &                                                 & 6 bins                      &  & -0.03$\pm$0.12   \\
\multicolumn{1}{l|}{}                                &  &                                                 & \textbf{10 bins}            &  & -0.02$\pm$0.13   \\
\multicolumn{1}{l|}{}                                &  &                                                 & 12 bins                     &  & -0.03$\pm$0.14   \\
\multicolumn{1}{l|}{}                                &  &                                                 &                             &  &                  \\
\multicolumn{1}{l|}{}                                &  & \textit{Initial Mass Function}                  & \textbf{Kroupa 2002}        &  & -0.02$\pm$0.13   \\
\multicolumn{1}{l|}{}                                &  &                                                 & Chabrier 2003               &  & -0.04$\pm$0.21    \\
\multicolumn{1}{l|}{}                                &  &                                                 &                             &  &                  \\
\multicolumn{1}{l|}{}                                &  & \textit{Noisy SED}                              & \textbf{3 \% uncertainties} &  & -0.02$\pm$0.13   \\
\multicolumn{1}{l|}{}                                &  &                                                 & Perturbed w/3\% unc.        &  & 0.15$\pm$0.72    \\ \hline
\multicolumn{1}{l|}{\multirow{5}{*}{Delayed-$\tau$}} &  & \multirow{2}{*}{\textit{Initial Mass Function}} & \textbf{Kroupa 2002}        &  & -0.39$\pm$0.12   \\
\multicolumn{1}{l|}{}                                &  &                                                 & Chabrier 2003               &  & -0.48$ \pm$0.89  \\
\multicolumn{1}{l|}{}                                &  &                                                 &                             &  &                  \\
\multicolumn{1}{l|}{}                                &  & \textit{Noisy SED}                              & \textbf{3 \% uncertainties} &  & -0.39$ \pm$0.12  \\
\multicolumn{1}{l|}{}                                &  &                                                 & Perturbed w/3\% unc.        &  & -0.47$\pm$1.16   \\ \hline
\end{tabular}
\caption{Summary statistics for SED models with differing parameter choices. Median stellar mass offsets are shown for each SED model distribution with the associated 1$\sigma$ width and the fiducial model choices are highlighted in bold. The choice of prior for the nonparametric model (Dirichlet+concentration parameter and continuity prior) have the largest effect on the stellar masses inferred from the nonparametric model. Using a perturbed input SED or a mismatched IMF slightly increase the median stellar mass offsets for each respective SFH model, though effectively each distribution is smeared out by the increased uncertainty in the inputs/model.  \textit{Concentration parameter}: The concentration parameter controls for the distribution of mass formation across bins. Low values ($\alpha$ $<$ 1)  prefer to put all of the weight in one bin while higher values more evenly ditribute the weight across all bins. The continuity prior is an alternative nonparametric model prior available in {\sc prospector} and favors smoother SFHs, explicitly weighting against sharp changes in mass formation between adjacent time bins. \textit{Time resolution}: The nonparametric model depends on a choice of time bins, both in number and location. Based on \cite{ocvirk_2006}, we set our time bins to be evenly spaced in logarithmic time and focus on the impact the time resolution (number of bins) has on the inferred stellar masses. \textit{IMF} and \textit{Noise}: For both the nonparametric and delayed-$\tau$ SFH model, we tested the impact of the assumed initial mass function (IMF) and the input SED noise level.}
\end{table}\label{table:appen_table}

\section{Variations in the SPS Model and Synthetic SEDs}\label{app:imf_noise}

\begin{figure}[h]
    \centering
    \includegraphics[width=0.48\textwidth]{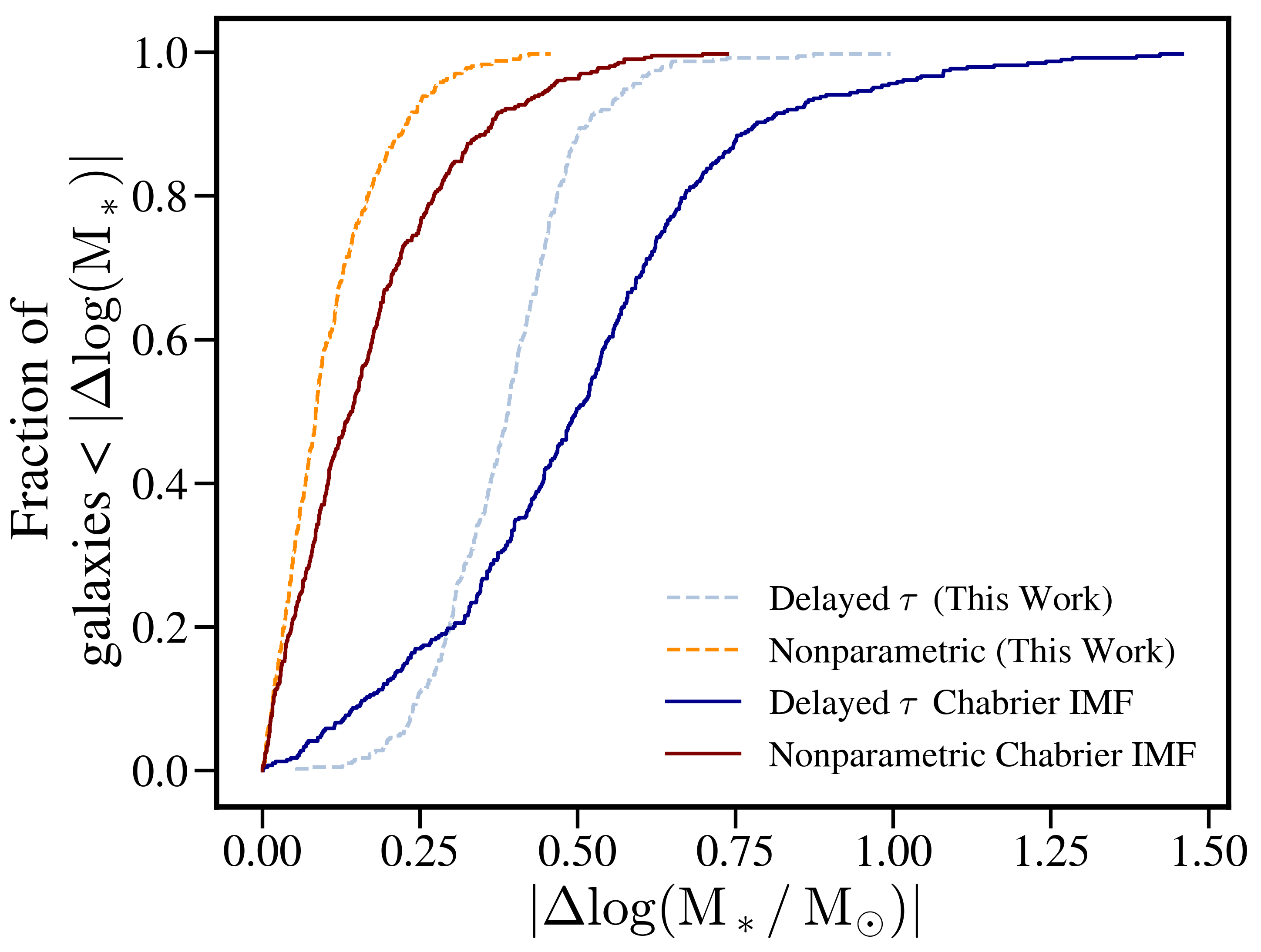}
    \includegraphics[width=0.48\textwidth]{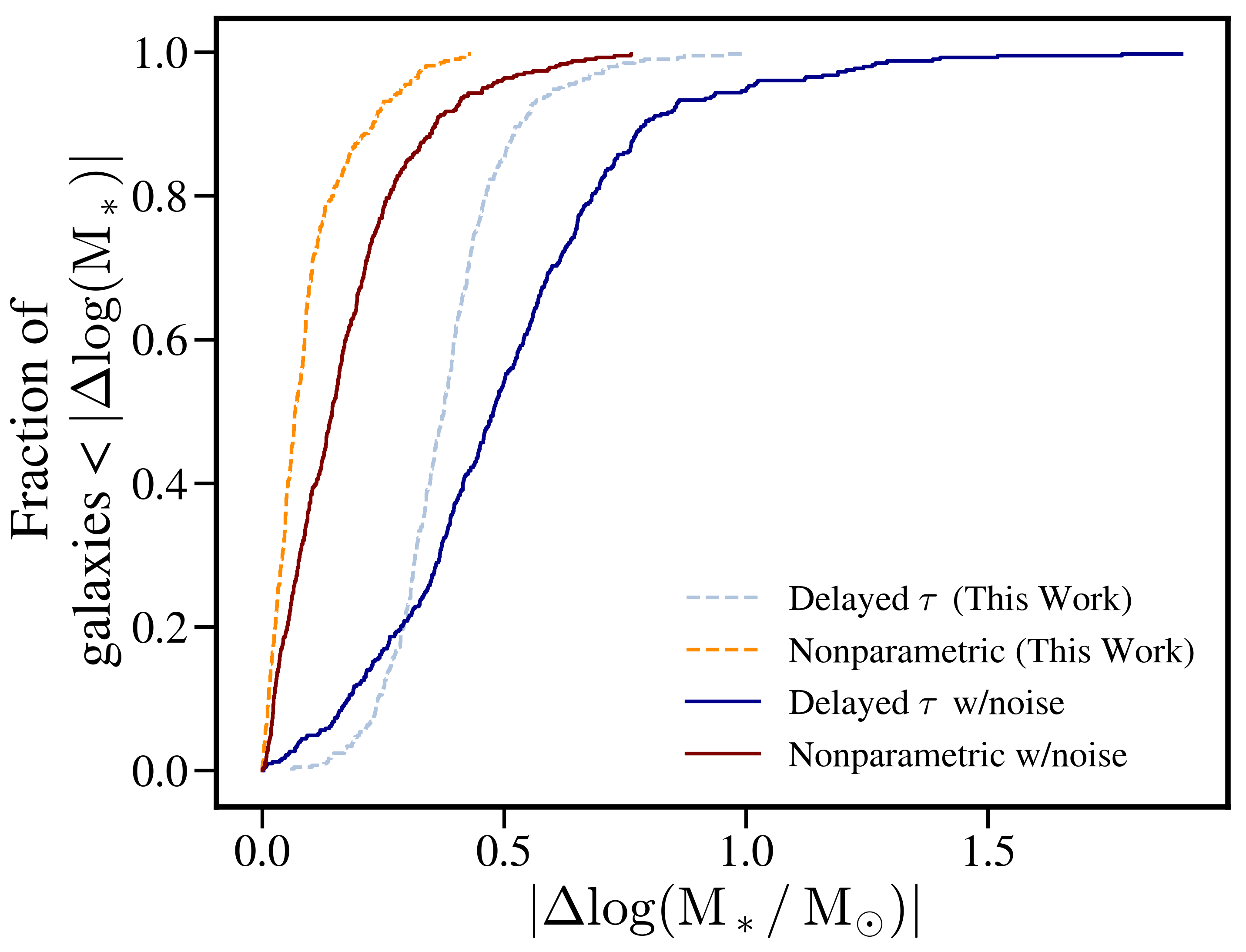}
    \caption{Distribution of stellar mass offsets for \textbf{Left}: differences in the input and model IMF and \textbf{Right}: SNR of the input broadband SED. Summary statistics are given for each data set and model.}
    \label{fig:appen_imf_noise}
\end{figure}

To quantify the impact of the assumed stellar population synthesis (SPS) model, we changed the model IMF in {\sc prospector} to no longer match the IMF used in the {\sc powderday} radiative transfer. Though the IMF represents just one uncertainty in an SPS model, the impact on the inferred stellar mass offsets is noticeable for both parametric and nonparametric SFH models, shown in the left panel of Figure \ref{fig:appen_imf_noise} with statistics given in Table \ref{table:appen_table}. Even so, the average galaxy has a more accurately inferred stellar mass from the nonparametric model than the delayed-$\tau$ model. An IMF change, to first order, is a multiplicative offset in stellar mass with no change to the observed photometry. Switching from a \cite{kroupa_initial_2002} IMF to a \cite{chab_imf} will decrease the mass by $0.03$ dex; fitting a \cite{salpeter_imf} IMF to photometry generated from a \cite{chab_imf} IMF would consistently over-estimate masses by $0.2$ dex.

Similarly, we simulated the effect of photometric noise on the inferred stellar mass. This was done by modulating the input photometry within the 3\% uncertainties. Again, we see noticeable decreases in accuracy from both nonparametric and parametric SFH models when noise is added, though the nonparametric SFHs continue to outperform the traditional parametric models, with offsets extending to $1.5$ dex.

\bibliography{bib2}{}
\bibliographystyle{aasjournal}

\end{document}